\documentclass[12pt]{article}

\usepackage[utf8]{inputenc}
\usepackage{graphicx}
\usepackage{fancyref}
\usepackage{amsmath}
\usepackage{latexsym}
\usepackage{xcolor}
\usepackage[]{units}
\usepackage{float}
\usepackage{physics}
\usepackage{mathrsfs}
\usepackage{amssymb}
\usepackage{enumerate}
\usepackage{parskip}
\usepackage[margin=1in]{geometry}
\usepackage{mathtools}
\usepackage{bm}
\usepackage{slashed}
\usepackage{tensor}
\usepackage{caption}
\usepackage{subcaption}
\RequirePackage[colorlinks=true, urlcolor=black, menucolor=black, linkcolor=purple, citecolor=blue]{hyperref}

\numberwithin{equation}{section}

\newcommand{\psibar}{\bar{\psi}}
\newcommand{\dslash}{\slashed{\nabla}}
\newcommand{\Dslash}{\slashed{D}}
\newcommand{\aslash}{\slashed{A}}

\def\tr{\,{\rm tr}\,}

\title{\textbf{Characters, quasinormal modes, and Schwinger pairs in $dS_2$ with flux}}
\author{\href{mailto:mg3978@columbia.edu}{Manvir Grewal} and 
\href{mailto:k.parmentier@columbia.edu}{Klaas Parmentier}\\\\ \it \small Center for Theoretical Physics, Department of Physics,\\ \it\small
Columbia University, New York, NY 10027, USA}
\date{}
\begin{document}
\fontseries{mx}\selectfont	
\maketitle
\begin{abstract}
An integral representation of the 1-loop partition function for charged scalars and spinors, minimally coupled to a uniform $U(1)$ field on $S^2$, is given in terms of $SO(1,2)$ Harish-Chandra group characters and evaluated exactly in terms of Hurwitz $\zeta$-functions. Analytically continuing the $U(1)$ field, we interpret the path integrals as quasicanonical partition functions in $dS_2$ with an electric field. The character itself is obtained as a trace over states living at the future boundary of de Sitter and has a quasinormal mode expansion.  The imaginary part of the partition function captures Schwinger pair creation in the static patch at finite temperature. The thermal enhancement is most noticeable for scalar masses below Hubble and leads to non-monotonicity of the current as a function of the field. This parameter range, when dimensionally reducing from a charged or rotating Nariai spacetime, is excluded by Swampland-inspired bounds. Around the $AdS_2$ black hole, in contrast to $dS_2$, there is a threshold to pair creation.
\end{abstract}
\thispagestyle{empty}
\newpage
{\hypersetup{linkcolor=black}
\tableofcontents
\thispagestyle{empty}}
\newpage
\addtocounter{page}{-2}
%%%%%%%%%%%%%%%%%%%%%%%%%%%%%%%%%%%%%
%%%%%%%%%%%%%%%%%%%%%%%%%%%%%%%%%%%%%
\section{Introduction and outline}\label{sec:intro}
%%%%%%%%%%%%%%%%%%%%%%%%%%%%%%%%%%%%%
%%%%%%%%%%%%%%%%%%%%%%%%%%%%%%%%%%%%%

Probing de Sitter ($dS$) space in a more precise and systematic way is not merely a perverse preoccupation of the perturbed physicist. On the contrary, such ventures can serve at least two important purposes. First of all, many models \cite{Maldacena:1998ih, Dong:2010pm, Coleman:2021nor} have been put forward over the years in an attempt to describe ($d$+1)-dimensional $dS$ space at a microscopic level, and in particular to clarify the meaning of the entropy associated to the cosmological horizon. In order to constrain or rule out such proposals, one needs precise macroscopic data to test them against. Secondly, in the Swampland spirit \cite{Vafa:2005ui}, many constraints have been proposed that aim to establish which low-energy effective field theories (EFT) can be consistently coupled to gravity. Some of the constraints relevant to $dS$ survive even in the limit where gravity is decoupled \cite{ Montero:2019ekk} and one could ask how they manifest within the EFT itself. 

As for the systematic extraction of exact macroscopic data, it was shown in \cite{Anninos:2020hfj} how to do so for arbitrary field content, by calculating 1-loop integrals on the sphere and interpreting them as quasicanonical thermal partition functions in the $dS$ static patch. The equivalence was demonstrated by making use of $SO(1,d+1)$ Harish-Chandra characters. The resulting character formalism has been developed for pure $dS$ in general dimension $d>1$. In the current work, our starting point is to include a uniform background electromagnetic field \eqref{eq: background} in the simplest case $d=1$. This gives a surprisingly rich toy model which links representation theory to vacuum instability and black hole superradiance. We determine the modifications required in the character formalism to capture the background and explore to what extent one can still extract exact static patch results from the sphere partition function. In particular, the electric field gives rise to pair creation, and we study how the thermal nature of the static patch affects this quantum instability of the vacuum. The $dS_2$ setup arises naturally through dimensional reduction from charged or rotating Nariai spacetimes, where recently proposed Swampland bounds \cite{Montero:2019ekk} constrain the EFT. Below, we give the outline of the paper and summarize our results. 

In Sec. \ref{sec: char}, we derive the character integral representation of the 1-loop partition function for charged scalars \eqref{eq:scalarchar} and spinors \eqref{eq:spinorchar} in the presence of the background Wu-Yang $U(1)$ field \eqref{eq: background} on $S^2$, starting from the Euclidean spectrum of the Klein-Gordon and Dirac operators \cite{Wu:1976ge}, as discussed in App. \ref{app:A}. The character formulas receive key modifications due to the background field. Encoded in them lies the quantum instability of the vacuum and these modifications conspire in such a way that the UV-divergences are left unchanged. At a technical level, the character representation allows one to simply read off these UV-divergences and algorithmically find the exact results \eqref{eq:scalar_fin} and \eqref{eq:spinor_fin} in terms of Hurwitz $\zeta$-functions. Taking the de Sitter length $l$ to be very large, the correct flat space results \cite{Blau:1988iz, Gavrilov:1996pz,Huet:2018ksz} are obtained . Following \cite{Anninos:2020hfj}, we calculate the static patch energy $U$ and entropy $S$ of the quantum fields. The background field makes the setup out-of-equilibrium, resulting in complex $U$ and $S$. At large values of the mass, the instability is sufficiently suppressed and it does become meaningful to discuss their real parts. Both increase monotonically with the field, except for the scalar at small values of the mass and field, as seen in Fig. \ref{fig:thermo} and \ref{fig:thermo2}.

In Sec. \ref{sec: algebra}, we write down the symmetry generators which commute with the equations of motion. We find that the background field shifts the standard $\mathfrak{so}(1,2)$ generators. Nonetheless, their action on boundary fields \eqref{eq: boundarygenerator} allows us to find the character $\chi(t)$ directly as the trace $\tr e^{-iHt}$. The quasicanonical interpretation of sphere partition functions stems from the fact that the character admits a quasinormal mode (QNM) expansion \cite{Anninos:2020hfj}. This is closely related to the fact that these QNMs can be constructed algebraically as a consequence of the large symmetry group of $dS$ \cite{Ng:2012xp, Jafferis:2013qia, Tanhayi:2014kba, Sun:2020sgn}. We verify that the same expansion of the character holds in our setup with background field. At the end of Sec. \ref{sec: algebra}, we also comment on Green functions and find the retarded propagator. Without flux, the conformal equivalence $dS_2 \times S^1 \sim AdS_3$ implies an $SL(2,\mathbb{R})\times SL(2,\mathbb{R})$ symmetry which factorizes the propagator \cite{Anninos:2011af}. This particular symmetry is broken by the presence of flux on $dS_2$, and by consequence, the structure of the retarded propagator in this case is more involved.  

The use of characters allows for an efficient extraction of the finite and imaginary parts of the 1-loop partition function. This imaginary part determines the persistence of the vacuum, which is subject to Schwinger pair production in the presence of a background electric field \cite{Schwinger:1951nm, Nikishov:1970br, Damour:1975pr,manogue}. In flat space, the relevant modes in a constant electric field were found long ago in \cite{ Narozhnyi:1976zs} and treated more recently in \cite{Akhmedov:2019rvx}. In Sec. \ref{sec: schwinger}, we calculate the scalar \eqref{eq:imscalar} and spinor \eqref{eq:imspinor} vacuum persistence for the $dS_2$ static patch by analytic continuation from $S^2$. The same method was applied for $AdS_2$ and $\mathbb{H}^2$ in \cite{Pioline:2005pf}. As in that case, the semiclassical instanton contributions come with alternating sign in the scalar case and same sign in the spinor case. Compared to $AdS_2$, however, in $dS_2$ there is no threshold for pair production. 

Our method of obtaining the Schwinger pair production rate is similar to the $\zeta$-function approach of \cite{Belgiorno:2009pq,Belgiorno:2009da}. The crucial difference is that we include the effect of the static patch thermal background by retaining periodicity in Euclidean time. Our setup is therefore more closely related to the work of \cite{Frob:2014zka}, where the current created in the $dS_2$ planar patch in the global vacuum was calculated. In particular, the semiclassical contributions \eqref{eq: semiclass} due the (anti)screening instantons \cite{Garriga:1993fh, Garriga:1994bm} appear with the same prefactor and relative sign in our result for the vacuum persistence as in the current found in \cite{Frob:2014zka}. We similarly find the same IR hyperconductivity, in contrast to the non-thermal static patch results \cite{Belgiorno:2009pq,Belgiorno:2009da}. In higher dimensions, the current was found in \cite{Bavarsad:2016cxh}. Pair production in $(A)dS$ has received much attention in the literature and has mostly been analyzed using Bogoliubov coefficients from a global perspective, including the cosmological pair production present without flux \cite{Kim:2008xv, Kim:2011hu, Cai:2014qba}, see also  \cite{Garriga:2012qp} for a discussion of the observer-dependence.  In the case without flux, propagators in various $dS$ patches, the imaginary part of the effective action, pair production, and the relation to rotation from Euclidean sphere have been discussed in  \cite{Akhmedov:2009ta, Akhmedov:2019esv}.

Furthermore, we apply the character formalism as developed for $AdS$ in \cite{Sun:2020ame} to study vacuum instability in the $AdS_2$ black hole background. Unlike in $dS_2$, there is a minimum value for the electric field above which pair creation becomes possible. Our results are in agreement with and clarify the $\zeta$-function regularized result \eqref{eq: imAdS} of \cite{Pioline:2005pf}. Finally, a physical setup in which charged particles in a uniform electric field on $dS_2$ appear is obtained by dimensionally reducing charged or rotating Nariai spacetimes \cite{Anninos:2010gh, Belgiorno:2009pq, Belgiorno:2009da}. We note in particular that the scalar mass region in which there is IR hyperconductivity \cite{Frob:2014zka} (see Fig. \ref{fig:imbehaviorscal}) is excluded by the Festina Lente (FL) bound \cite{Montero:2019ekk, Montero:2021otb}. Pair creation, which does not depend on the details of the gravitational collapse \cite{Birrell:1982ix}, is discussed more generally for de Sitter black holes in \cite{Bavarsad:2016cxh, Chen:2020mqs}. The $AdS_2$ setup arises in the near-horizon extremal Kerr limit \cite{Anninos:2019oka}.

%%%%%%%%%%%%%%%%%%%%%%%%%%%%%%%%%%%%%
%%%%%%%%%%%%%%%%%%%%%%%%%%%%%%%%%%%%%
\section{Character integrals for particles in background flux}\label{sec: char}
%%%%%%%%%%%%%%%%%%%%%%%%%%%%%%%%%%%%%
%%%%%%%%%%%%%%%%%%%%%%%%%%%%%%%%%%%%%
In this section, we will rewrite the 1-loop partition function for charged particles on a sphere with background $U(1)$ flux in terms of a character integral. There are several advantages of doing so. At a technical level, this allows one to simply read off the UV-divergences and algorithmically find the exact result in terms of Hurwitz $\zeta$-functions. At a conceptual level, the result can moreover be interpreted as the quasicanonical partition function in the de Sitter static patch, as shown in \cite{Anninos:2020hfj}. In this section, we will apply these methods to $dS_2$ and generalize them by the inclusion of a background uniform $U(1)$ field, to which we minimally couple charged scalars and spinors. 

Our starting point will thus be to Wick rotate the $dS_2$ static patch to $S^2$ of radius $l$
\begin{equation}
	ds^2 = l^2(d\theta ^2 + \sin^2 \theta d\varphi^2) \;.
\end{equation}
For most of this paper, we will set $l=1$, restoring it when needed by dimensional analysis. The 2-sphere can be covered by the usual two charts
\begin{equation}\begin{split}
	U_N &\equiv \{ (\theta, \varphi) | \ 0 \le \theta \le \frac{\pi}{2} + \epsilon \}  \,, \\
	U_S & \equiv \{ (\theta, \varphi) | \ \frac{\pi}{2} - \epsilon \le \theta \le \pi \}  \,,
\end{split}\end{equation}
on which we will consider the Wu-Yang $U(1)$-connection \cite{Wu:1976ge}
\begin{equation}\begin{split}\label{eq:wu}
	A_N &= \hspace{9pt} B(1-\cos \theta) d\varphi  \,, \\
	A_S &=-B(1+\cos \theta) d\varphi \,,
\end{split}\end{equation}
corresponding to the potential produced by a magnetic monopole located at the center of the sphere. For the gauge transformation on $U_N \cap U_S$ to be well-defined, one needs $2B \in \mathbb{Z}$. The field strength is then determined to be proportional to the volume form
\begin{equation}\label{eq: background}
	F = dA = B \sin\theta\;  d\theta \wedge d\varphi \,,
\end{equation}
and clearly keeps the $SO(3)$ symmetry. In the $dS_2$ static patch, the analytic continuation $B \to iE$ corresponds to having a constant electric field. For simplicity of notation, we will assume that the field $B>0$ and set the charge $e=1$.

%%%%%%%%%%%%%%%%%%%%%%%%%%%%%%%%%%%%%
\subsection{1-loop partition function in the scalar case}\label{sec: scal1loop}
%%%%%%%%%%%%%%%%%%%%%%%%%%%%%%%%%%%%%
We wish to evaluate the Euclidean path integral for a charged scalar on $S^2$, minimally coupled to a $U(1)$ gauge field with uniform field strength,
\begin{equation}
%	Z_{PI} = \int \mathcal{D}\phi^* \mathcal{D}\phi \ e^{-\int \phi^* [-(\vec{\nabla}-i\vec{A})^2 + m^2] \phi} \,.
	Z_{\phi} = \int \mathcal{D}\phi^* \mathcal{D}\phi \ e^{-\int \phi^* (D^2 + m^2) \phi} \,,
	\label{eq:scalarpartition}
\end{equation}
where $D^2 \equiv - (\vec{\nabla} - i \vec{A})^2$. A convenient UV-regularized version of this path integral is obtained via the heat kernel method \cite{Vassilevich_2003}:
\begin{equation}
	\log Z_{\phi,\epsilon} = \int_0^{\infty} \frac{d\tau}{\tau} \ e^{-\epsilon^2 / 4 \tau}  \Tr e^{-\tau(D^2 +m^2)} \,.
\end{equation}

\begin{figure}[ht]
    \centering
   \begin{subfigure}{0.3\textwidth}
            \centering
            \includegraphics[width=\textwidth]{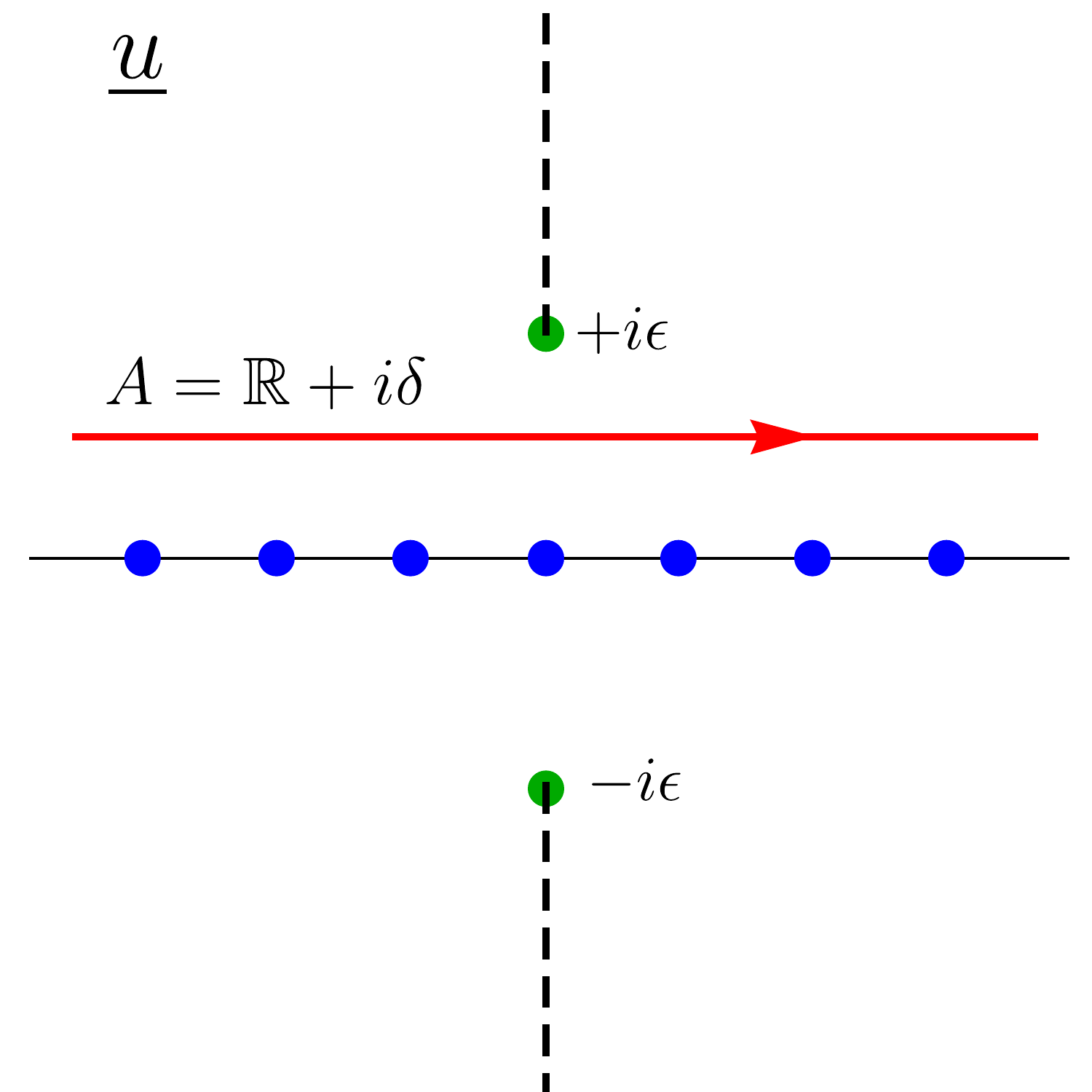}
            \caption[]%
            {{\small original contour}
             }    
        \end{subfigure}
        \begin{subfigure}{0.3\textwidth}  
            \centering 
            \includegraphics[width=\textwidth]{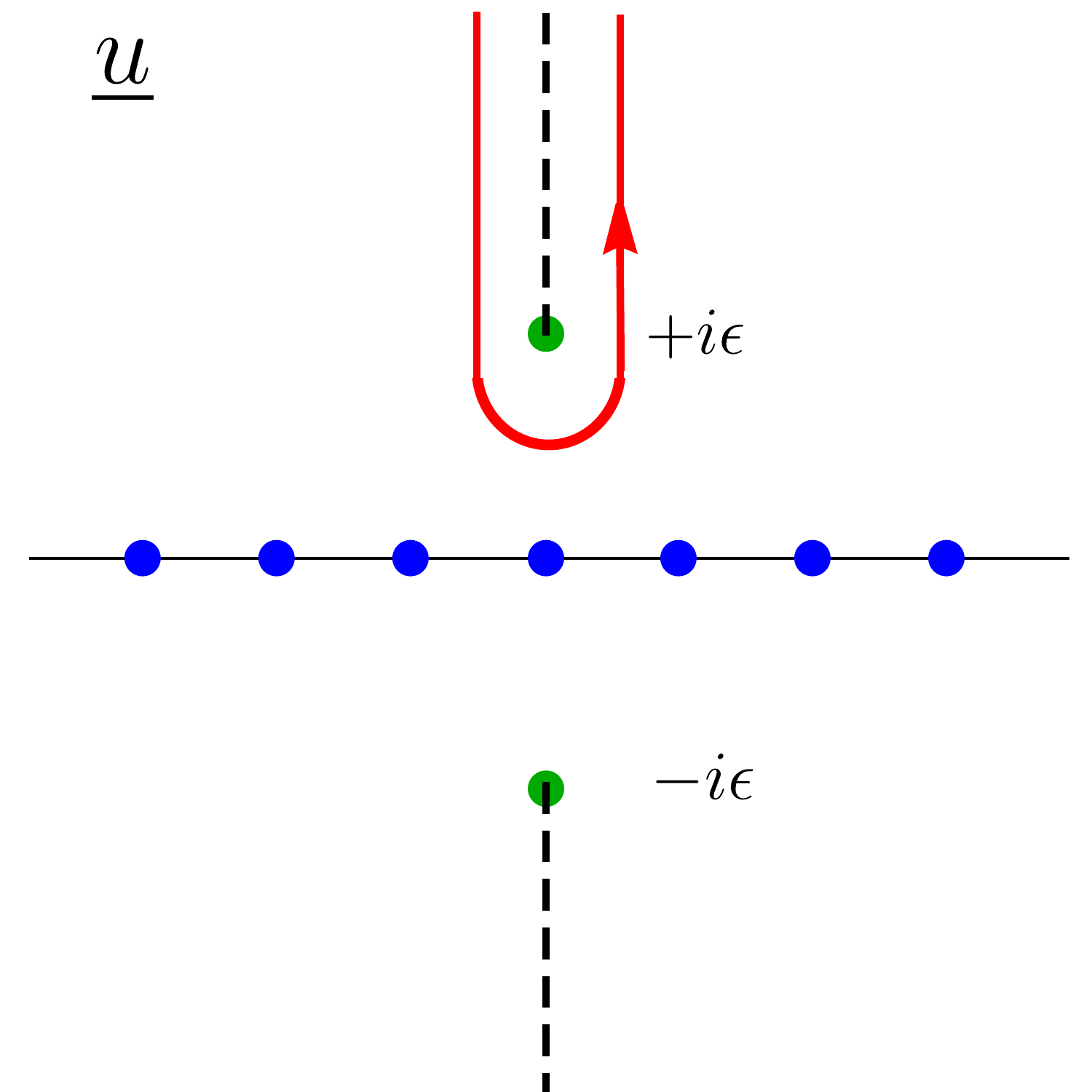}
            \caption[]%
            {{\small \centering  folded contour}}   
        \end{subfigure}
        \begin{subfigure}{0.3\textwidth}  
            \centering 
            \includegraphics[angle=270,width=\textwidth]{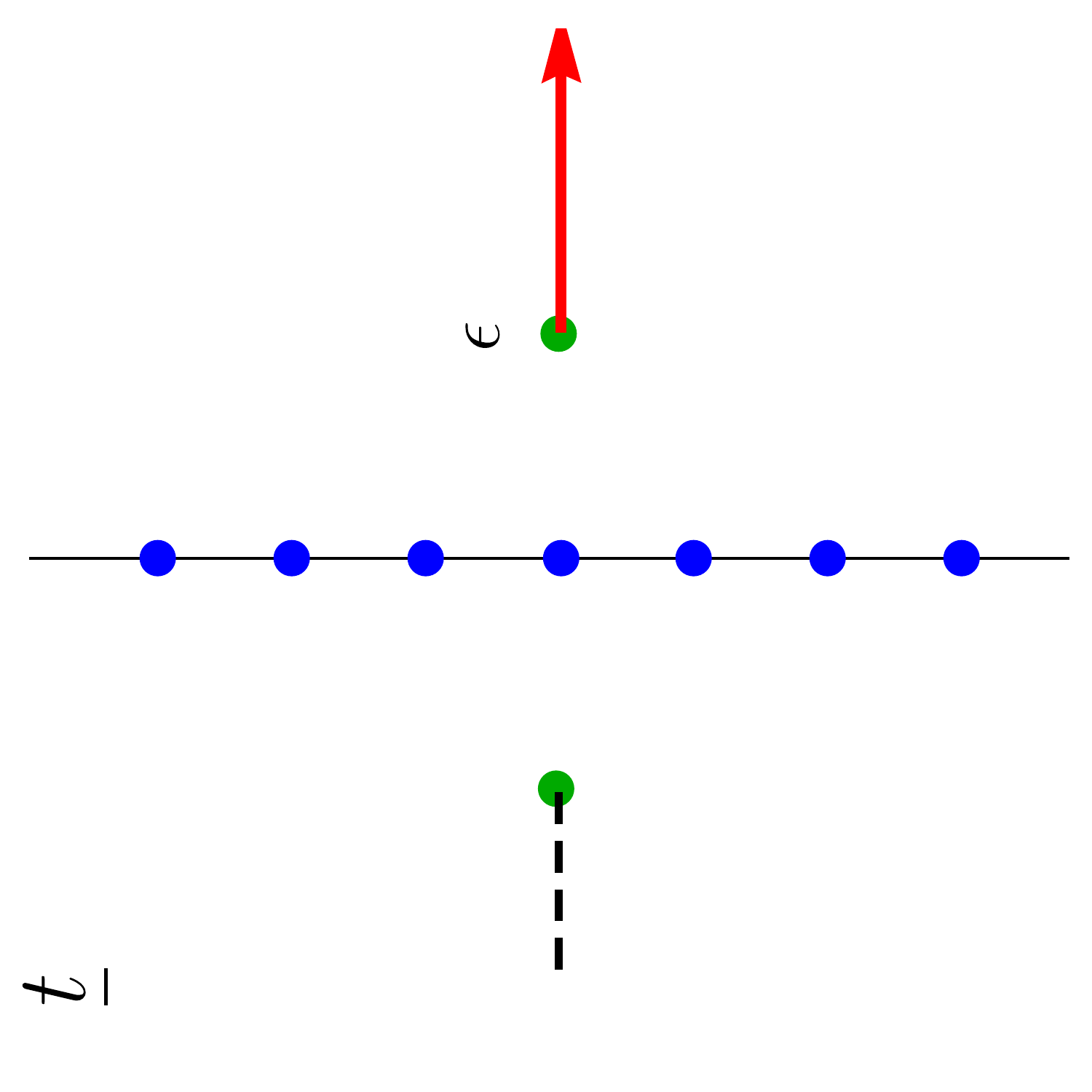}
            \caption[]%
            {{\small \centering  rotated folded contour}}   
        \end{subfigure}
    \caption{To arrive at the character integral from the Hubbard-Stratonovich trick, we fold the contour $A$ (red) along the branch cut from \eqref{hubbardeq} with branch point $+i \epsilon$ (green dot), and then rotate $u=it$. The blue dots represent the poles of $f(u)$. }\label{fig:hubbard}
\end{figure}

Using the spectrum of $D^2$, as reviewed in App. \ref{app:A1}, we find
\begin{equation}
	\log Z_{\phi,\epsilon} = \int_0^{\infty} \frac{d\tau}{\tau} \ e^{-\epsilon^2 / 4\tau} e^{-\tau\eta_\phi^2} \sum_{n=0}^{\infty} D_{n+B}^{3} e^{-\tau(n+B+\frac{1}{2})^2} \,, \qquad \eta_\phi \equiv \sqrt{m^2-B^2 - \frac{1}{4}} \,,
	\label{eq:logZtauintegral}
\end{equation}
where $D_{n + B}^3$ is the degeneracy of the $n$th eigenvalue of $D^2$. Following \cite{Anninos:2020hfj}, we use the Hubbard-Stratonovich trick to perform the sum over $n$, writing
\begin{equation}
	\sum_{n=0}^\infty D_{n+B}^{3} e^{-\tau(n+B+\frac{1}{2})^2} = \int_A du \ \frac{e^{-u^2/4\tau}}{\sqrt{4\pi \tau}} f(u) \,, \quad f(u) \equiv \sum_{n=0}^\infty D_{n+ B}^{3} e^{iu(n+B+\frac{1}{2})} \,,
\end{equation}
with integration contour $A = \mathbb{R} + i\delta$, $\delta >0$ (see Fig. \ref{fig:hubbard}). Using the degeneracies \eqref{appA1:deg}, the sum can now be evaluated to give
\begin{equation}\begin{split}
	f(u) &= \sum_{n=0}^\infty \Big(2\Big(n+B\Big)+1\Big) e^{iu(n+B+\frac{1}{2})}  \\
	%& = e^{iu (B+1/2)} \sum_{n=0}^\infty (D_n^{3} +2B )e^{iun}  \\
	&= \Big(\frac{1+e^{iu}}{1-e^{iu}} +2B \Big) \frac{e^{iu(B+1/2)}}{1-e^{iu}} \,.
\end{split}\end{equation}

Consider first real, positive $\eta_\phi$, i.e. $4m^2> 4B^2+1$, corresponding to the complementary series representations of $SO(1,2)$. Keeping $\Im u = \delta <\epsilon$, we can now perform the $\tau$-integral to obtain 
\begin{equation} \label{hubbardeq}
	\log Z_{\phi,\epsilon} = \int_A \frac{du}{\sqrt{u^2+\epsilon^2}} \ e^{-\eta_\phi\sqrt{u^2+\epsilon^2}}f(u) \,.
\end{equation}
After deforming the contour as in Fig. \ref{fig:hubbard} to wrap along the branch cut and changing variables to $u = it$, the integral becomes
\begin{equation}
	\log Z_{\phi,\epsilon} = \int_{\epsilon}^{\infty} \frac{dt}{\sqrt{t^2-\epsilon^2}} \ \Big(\frac{1+e^{-t}}{1-e^{-t}} +2B \Big) \frac{e^{-(B+\frac{1}{2})t + i\eta_\phi \sqrt{t^2-\epsilon^2}} + e^{-(B+\frac{1}{2})t - i\eta_\phi \sqrt{t^2-\epsilon^2}}}{1-e^{-t}} \,.
\end{equation}

The result for imaginary $\eta_\phi$, corresponding to the principal series representations, is obtained by analytic continuation. Formally taking the $\epsilon \rightarrow 0$ limit above, we obtain the character integral
\begin{equation}\label{eq:scalarchar}
\boxed{	\log Z_{\phi} = \int_0^\infty \frac{dt}{t} \ \Big(\frac{1+e^{-t}}{1-e^{-t}} + 2B\Big) e^{-Bt}\chi(t) \,, \quad \chi(t) \equiv \frac{e^{-(\frac{1}{2}+i\eta_\phi)t } + e^{-(\frac{1}{2}-i\eta_\phi)t}}{1-e^{-t}} \,. \,}
\end{equation}
Here $\chi(t)$ is the $SO(1,2)$ character of the unitary irreducible representation characterized by $\Delta = \frac{1}{2} + i \eta_\phi$. Looking at \eqref{eq:scalarchar}, we note that, apart from the appearance of $B$ in $\eta_\phi$, there are two more key modifications compared to the case $B=0$, which was discussed in \cite{Anninos:2020hfj}. Indeed, the background field appears both as an extra term in the prefactor and through an exponential factor. The former stems from the Landau levels in flat space, as can be seen by taking the $dS$ length very large. The other term in the prefactor, $(1+e^{^t})(1-e^{-t})^{-1}$ is also present when $B=0$ and takes into account the compact geometry. On the other hand, the exponential factor $e^{-Bt}$ contributes as a shift in energy.

UV-divergences arise when $\epsilon$, or equivalently $t$, becomes small. By locality, there should not be any $\epsilon^{-1}$ divergence of the partition function for odd $d$. This includes our situation, namely $d=1$. For $B=0$, this was ensured by the fact that the integrand in \eqref{eq:scalarchar} is odd in $t$. For general $B$, this is not obvious at first sight since the integrand no longer has a definite parity.  Nonetheless, the UV (small-$t$) expansion gives
\begin{equation}\label{eq:uvscal}
    \frac{1}{t} \Big(\frac{1+e^{-t}}{1-e^{-t}} + 2B\Big)e^{-B t} \chi(t) = 2\Big(\frac{2}{t^3} + \frac{ \frac{1}{3} - m^2}{t} + \ldots \Big) \,.
\end{equation}
The absence of a $t^{-2}$ term means that there is no divergence of the form $\epsilon^{-1}$. Moreover, the divergences that do arise are independent of the flux. In general, the leading divergence of the small-t expansion
has a coefficient of $2n$, where $n$ is the number of on-shell degrees of freedom, which for a complex scalar is $n=2$. The $t^{-1}$ term is the Weyl anomaly and has coefficient $\frac{c}{3}$ when $m=0$, where $c$ is the central charge. In this case, it is that of 2 real free bosons, namely $c=2$.

Using the above UV-expansion and the heat kernel regularization scheme of App.~C of \cite{Anninos:2020hfj}, we can then exactly evaluate \eqref{eq:scalarchar}, leading to the final result
\begin{equation}\begin{split} \label{eq:scalar_fin}
\log Z_\phi = \; &   2\zeta'(-1,\Delta+B)+2 \zeta'(-1,\bar{\Delta}+B)-2 i\eta_\phi\zeta'(0,\Delta+B)\\&+ 2i\eta_\phi\zeta'(0,\bar{\Delta}+B) +  2\eta_\phi^2 +  \Big( \frac{2}{3} - 2 m^2 \Big) \log M   
	+ \frac{4}{\epsilon^2} \,,
\end{split}\end{equation}
where $M \equiv 2e^{-\gamma}/\epsilon$ and $\gamma$ is the Euler-Mascheroni constant. In Sec. $\ref{sec: thermo}$ and $\ref{sec: schwingscal}$ respectively, we will further analyze and interpret the real and imaginary parts of the partition function that arise when continuing to imaginary flux.

Finally, let us comment on the flat space limit. To restore factors of $l$, we must restore $l$ in the metric and take $D^2 \rightarrow D^2 / l^2$ in \eqref{eq:scalarpartition}. Keeping track of such factors throughout the above derivation then leads to the modification $m \rightarrow m l$ and $B\rightarrow B l^2$. In Lorentzian signature we rotate $B \to iE$. The flat space limit is obtained by taking $l \to \infty$, while keeping $m, E$ fixed in \eqref{eq:scalarchar}. Upon subtracting the background energy, we then retrieve the flat space result \cite{Blau:1988iz, Huet:2018ksz}
\begin{equation}\label{eq:flatscal}
    \log Z_{\phi,\text{flat}} =   \frac{i E V}{4\pi} \int \frac{d \tau}{\tau} e^{i \tau \frac{m^2}{E}} \Big(\frac{1}{\sinh \tau}-\frac{1}{\tau} \Big) \,,
\end{equation}
with $V= 4\pi l^2$ being the Euclidean volume. This can be seen in several ways. At the level of the character integral, one can for instance use the Riemann-Lebesgue lemma to see that only terms with $\eta_\phi- E$ survive. Perhaps the most insightful way, however, starts at the level of the spectrum (\ref{appA1:eig}-\ref{appA1:deg}), which at large $B$ describes the non-relativistic Landau levels. 
% From this point of view, the $i E l^2$ in the prefactor of \eqref{eq:scalarchar} simply captures the Euler-Heisenberg part of the effective action, while the other term $(1+e^{-t})(1-e^{-t})^{-1}$ captures the thermal nature, due to the periodicity on $S^2$ in the angle $\varphi$, corresponding to the Euclidean time of the $dS_2$ static patch.

%%%%%%%%%%%%%%%%%%%%%%%%%%%%%%%%%%%%%
\subsection{1-loop partition function in the spinor case}
%%%%%%%%%%%%%%%%%%%%%%%%%%%%%%%%%%%%%
In this section, we wish to evaluate the path integral for a minimally coupled Dirac spinor
\begin{equation}
	Z_{\psi} = \int \mathcal{D} \psibar \mathcal{D}\psi \ e^{-\int \psibar [(\dslash -i\aslash) + m] \psi} \,,
\end{equation}
%We regularize this fermionic path integral as
%\begin{equation}
%	\log Z_{\psi,\epsilon} = -\int_0^{\infty}  \frac{d\tau}{\tau} \ e^{-\epsilon^2 / 4 \tau}  \Tr  e^{-\tau(i \Dslash +m)} \,,
%\end{equation}
where $\Dslash \equiv -i(\dslash - i \aslash)$, following essentially the same steps as in the scalar case.

The spectrum of $\Dslash$ is obtained in App.~\ref{app:A2}.
Note that the non-zero modes of $\Dslash$ come in pairs $\lambda_{N,n}= \pm \sqrt{n(n+2B)}$ whereas the zeromodes ($n=0$) are unpaired. Pairing the non-zero eigenvalues together in the heat kernel regularization, we may rewrite the path integral as
\begin{equation}
	\log Z_{\psi,\epsilon} = - \int_0^{\infty} \frac{d\tau}{\tau} \ e^{-\epsilon^2 / 4\tau} e^{-\tau \eta_\psi^2} \Big(\sum_{n=1}^{\infty} D_{n +B-\frac{1}{2},\frac{1}{2}}^{3} e^{-\tau(n+B)^2} + B e^{-\tau B^2}\Big)\,, \end{equation}
where
\begin{equation}
\eta_\psi \equiv \sqrt{m^2 - B^2} \,.
\end{equation}
The first term in parentheses stems from states which in the flat space limit correspond to the two towers of Landau levels, one for each spin, while the second term corresponds to the unpaired zeromodes. Now we apply the Hubbard-Stratonovich trick and integrate along the same contour as in the scalar case. After formally taking $\epsilon \rightarrow 0$, the integral becomes
\begin{equation}\boxed{
   \log Z_{\psi}= -\int^\infty_0 \frac{dt}{t}\Big(\csch{\frac{t}{2}} + 2B \cosh{\frac{t}{2}}\Big)e^{-B t}\chi(t)\;, \;\; \chi(t) \equiv \frac{e^{-(\frac12 +i\eta_\psi) t}+e^{-(\frac12 -i\eta_\psi) t}}{1-e^{-t}} \,. }\label{eq:spinorchar}
\end{equation}
Compared to the case $B=0$, we see that the structural modifications to the character integral are parallel to what we discussed below \eqref{eq:scalarchar} for the scalar result.

As a check of locality, there should not be any $\epsilon^{-1}$ divergence of the partition function. Indeed we see
\begin{equation}\label{eq:uvspin}
\frac{1}{t}\Big(\csch{\frac{t}{2}} + 2B \cosh{\frac{t}{2}} \Big)e^{-B t}\chi(t) = 2\Big( \frac{2}{t^3} - \frac{ \frac{1}{6} + m^2}{t} + \ldots \Big) \,,
\end{equation}
implying the absence of a UV-divergence of the form $t^{-2}$ which would give an $\epsilon^{-1}$ divergence. The coefficients in \eqref{eq:uvspin} are understood from the on-shell degrees of freedom and central charge of a Dirac spinor, $n=2$ and $c=1$, corresponding to 2 free fermions. That the $\epsilon^{-2}$ deep-UV-divergence does not depend on $B$ is clear. For the logarithmic Weyl anomaly term, note that in the massless case, left- and right-moving fermions $\psi_\pm$ in $d=1$ can be understood as neutral fermions propagating in a background with a complexified Weyl factor \cite{Anninos:2019oka}. For the sphere with uniform field $B$, it takes the form $(1\pm 2i B)\log\sech u$, with $u= \log \tan\frac{\theta}{2}$ in terms of the usual $S^2$-coordinate. In general, such complexified Weyl transformation gives rise to an anomaly $\frac{c}{6}(R\mp 4i B)$. Combining now the contributions from $\psi_{\pm}$, one finds that the field strength cancels out and the coefficient of the $t^{-1}$ term is given by $\frac{c}{3}$ in the massless case, as argued before.

We can now compare \eqref{eq:scalar_fin} and \eqref{eq:spinor_fin} and try to choose the field content such that the UV-divergences cancel and the partition function is finite. The leading $\epsilon^{-2}$ is cancelled for any value of the mass by including the same amount of Dirac spinors as complex scalars. However, also demanding cancellation of the $\log\epsilon$ divergences imposes a nontrivial relation between the masses of the two fields. From \eqref{eq:scalar_fin} and \eqref{eq:spinor_fin}, we find that the result will be UV-finite if $m_\phi^2 = m_\psi^2 + \frac{1}{2}$, or equivalently $\eta_\phi^2 = \eta_\psi^2 + \frac{1}{4}$. 

In heat kernel regularization, the spinor character formula \eqref{eq:spinorchar} can be exactly evaluated. We find 
\begin{equation}\begin{split} \label{eq:spinor_fin}
	\log Z_\psi = & -2 \zeta'\Big(-1,\Delta + B+\frac12\Big)-2\zeta'\Big(-1,\bar{\Delta} + B+\frac12\Big) +2 i\eta_\psi\zeta'\Big(0,\Delta+B+\frac12\Big)\\ &-2i\eta_\psi\zeta'\Big(0,\bar{\Delta}+B+\frac12\Big)  
	 -2  \eta_\psi^2+  \Big( \frac{1}{3} + 2m^2 \Big) \log M - \frac{4}{\epsilon^2}\,,
\end{split}\end{equation}
where $M\equiv 2e^{-\gamma}/\epsilon $ as before.

As an example, let us consider the special case of a massless spinor. In the massless limit $\eta_\psi = i B$, \eqref{eq:spinor_fin} simplifies. In particular, the Hurwitz $\zeta$-functions  satisfy \cite{dlmf}
\begin{equation}\begin{split}
	\zeta'(0,a) &= \log  \frac{\Gamma(a)}{\sqrt{2\pi}} \,, \\
	\zeta'(-1,2B+1) &= \zeta'(-1) + \log(H(2B)) \,,
\end{split}\end{equation}
where $\zeta'(-1) = \frac{1}{12} - \log A$, with $A \approx 1.282$ being the Glaisher-Kinkelin constant. $H$ is the hyperfactorial function, which for integer arguments, such as $2B$ on the sphere, takes the form
\begin{equation}
	H(2B) = \prod_{k=1}^{2B} k^k \,,
\end{equation}
The finite part of \eqref{eq:spinor_fin} then becomes
\begin{equation}\begin{split}
	Z_{\psi,2B}^\text{fin} &= e^{-4 \zeta'(-1)}\frac{[\Gamma(2B+1)]^{2B}}{H(2B)^2}e^{2B^2} \,.
\end{split}\end{equation}

Finally, in the flat space limit, taking into account the background subtraction, the path integral correctly gives \cite{Blau:1988iz,Huet:2018ksz}
\begin{equation}\label{eq: flatspin}
    \log Z_{\psi, \text{flat}} = - \frac{iEV}{4\pi} \int^\infty_0 \frac{d\tau}{\tau}e^{i \tau \frac{m^2}{E}}\Big(\coth{\tau}-\frac{1}{\tau}\Big) \,,
\end{equation}
with $V= 4\pi l^2$ being the Euclidean volume. This reduction is most easily seen at the level of the spectrum. The coth term essentially captures the contribution from the towers of Landau levels built on the two spin states. The other term in \eqref{eq:spinorchar} captures the thermal nature of the sphere partition when interpreted in the static patch, as we will review in the next section. 

%%%%%%%%%%%%%%%%%%%%%%%%%%%%%%%%%%%%%
\subsection{Static patch thermodynamics}\label{sec: thermo}
%%%%%%%%%%%%%%%%%%%%%%%%%%%%%%%%%%%%%
When we turn off the flux in our 1-loop calculations, we are effectively dealing with a free QFT. Bosonic and fermionic oscillator modes of frequency $\omega$ are in thermal equilibrium in the static patch, at inverse temperature $\beta$. Using an integral representation of the logarithm, one can represent the thermal partition function as \cite{Anninos:2020hfj}
\begin{equation}\label{eq: thermal}
    \log \Tr e^{-\beta H} = \int^\infty_0 \frac{dt}{2t} \Big(\frac{1+e^{-2\pi t/\beta}}{1-e^{-2\pi t/\beta}}\chi_{\text{bos}}(t) - \frac{e^{-2\pi t/\beta}}{1-e^{-2\pi t/\beta}} \chi_{\text{fer}}(t)\Big)  =\log Z \,,
\end{equation}
where the Harish-Chandra character is the Fourier transform of the density of states\footnote{More precisely, we are interested in the trace over static patch excitations  whereas the character is calculated from the density of states in the global vacuum. There is however a one-to-one map between them, as detailed in \cite{Anninos:2020hfj}. In Sec. \ref{sec: bound}, we will calculate the character directly from the action of the $\mathfrak{so}(1,2)$ generators.}
\begin{equation}
    \chi(t) = \int^\infty_{-\infty}\rho(\omega) e^{i\omega t} \,.
\end{equation}
The relative factor of 2 difference between our results at zero flux and \eqref{eq: thermal} is because the latter was derived for uncharged particles. One can now use $\beta = 2\pi l$ to obtain thermodynamical quantities of interest, such as the energy $U$ or entropy $S$, by taking suitable derivatives of the partition function
\begin{equation}\label{eq:thermalUS}
    U = -2\pi\partial_l \log Z \,, \quad S = (1-l\partial_l)\log Z \,.
\end{equation}
Several examples were given in \cite{Anninos:2020hfj}. The entropy can be interpreted as an entanglement entropy across the static patch horizon. Its leading divergence follows an area law, whereas the coefficient of the logarithmic divergence is universal.

In the presence of a background flux, the setup becomes inherently time-dependent and the system is no longer in equilibrium. Even worse, the vacuum is rendered unstable due to the spontaneous creation of Schwinger pairs. As we shall discuss in detail in Sec. \ref{sec: schwinger}, this effect manifests itself through an imaginary contribution to $\log Z$. By consequence, if one chooses to apply the equilibrium expressions \eqref{eq:thermalUS}, one would find complex $U$ and $S$. For the energy, the imaginary part has a familiar interpretation as a decay width, whereas it is a priori not clear what to make of the entropy in this case. Nevertheless one can adopt a pragmatic point of view and simply study its real part. In particular, when the particle mass $m$ is large or the field value $E$ is small, pair creation effects are quite suppressed and it should be meaningful in this regime to study thermodynamic properties of the system.  

\begin{figure}[ht]
    \centering
   \begin{subfigure}{0.4\textwidth}
            \centering
            \includegraphics[width=\textwidth]{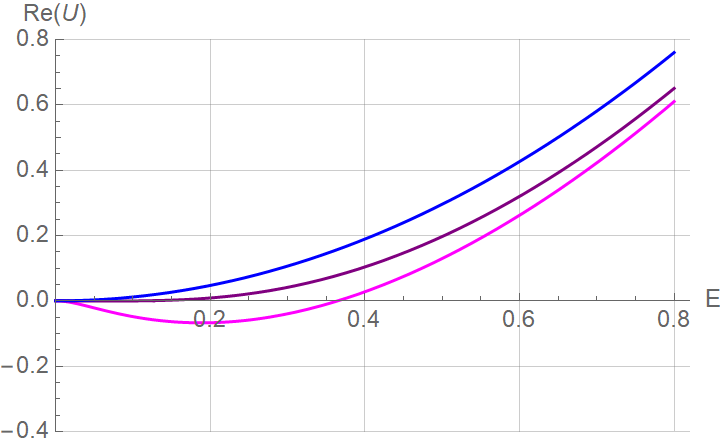}
            \caption[]%
            {{\small energy as function of electric field}
             }    
        \end{subfigure}
        \hspace{0.5cm}
        \begin{subfigure}{0.385\textwidth}  
            \centering 
            \includegraphics[width=\textwidth]{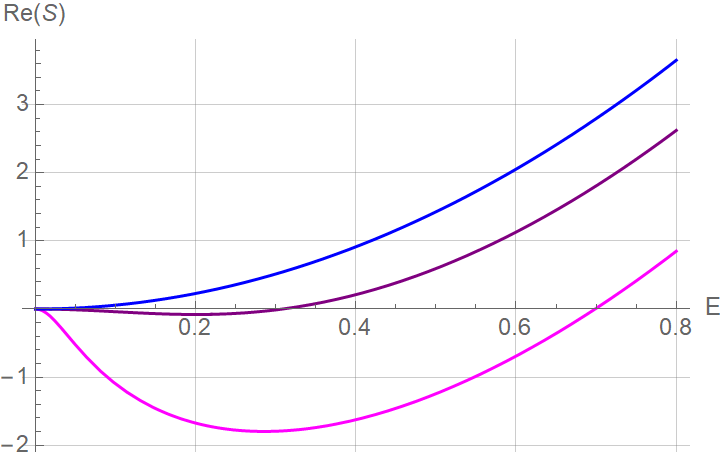}
            \caption[]%
            {{\small \centering  entropy as function of  electric field}}    
        \end{subfigure}
    \caption{When turning on a flux in the static patch, the Euclidean partition function of the charged scalar becomes complex. Using \eqref{eq:thermalUS} we calculate the finite parts of the energy $U$ and entropy $S$ as function of the electric field $E$. In the above plots, we display the real parts, where the value at zero field has been subtracted. From magenta to blue we have $m=0.2,\; 0.5,\; 1$. For sufficiently large mass, these quantities are monotonic as function of $E$. For smaller masses $4 m^2 \lesssim 1$ this is not the case. In Sec. \ref{sec: schwinger}, we will find a similar effect in the Schwinger pair production rate. The low mass range is excluded by the FL bound \cite{Montero:2021otb}.}\label{fig:thermo}
\end{figure}

\begin{figure}[ht]
    \centering
   \begin{subfigure}{0.4\textwidth}
            \centering
            \includegraphics[width=\textwidth]{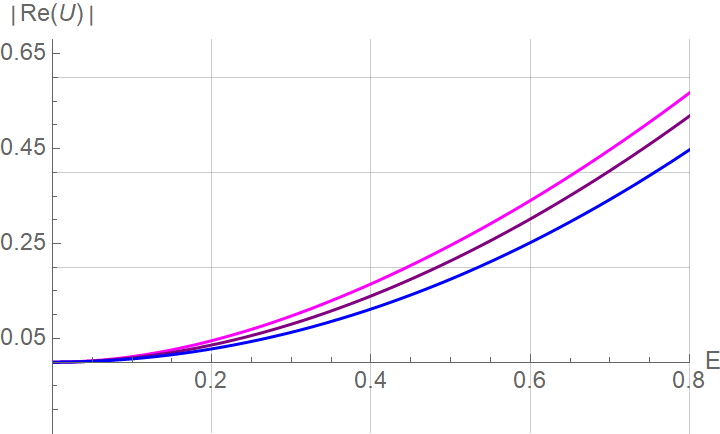}
            \caption[]%
            {{\small energy as function of electric field}}    
        \end{subfigure}
        \hspace{0.5cm}
        \begin{subfigure}{0.39\textwidth}  
            \centering 
            \includegraphics[width=\textwidth]{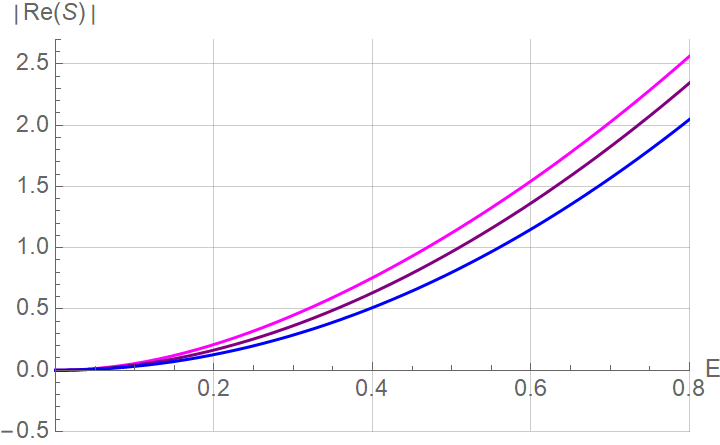}
            \caption[]%
            {{\small entropy as function of  electric field}}    
        \end{subfigure}
    \caption{For the charged spinor, we calculate the finite parts of the energy $U$ and entropy $S$ as function of the electric field $E$ as we did for the scalar in Fig. \ref{fig:thermo}. The main difference is that for spinors, $U$ and $S$ are both monotonic as function of electric field $E$, which was not the case for scalars. From magenta to blue we have $m=0.2,\; 0.5,\; 1$.}\label{fig:thermo2}
\end{figure}

The Hartle-Hawking state prepared by the path integral on the hemisphere depends on the electric field, and therefore the static patch energy and entropy will too. The $dS$ length is reinstated in \eqref{eq:scalarchar} and \eqref{eq:spinorchar} by $m \rightarrow m l$ and $B \to i E l^2$, after which we can apply \eqref{eq:thermalUS}. From the UV expansions \eqref{eq:uvscal} and \eqref{eq:uvspin}, we reach the same conclusions for the divergent parts of the entropy as in the case without flux: the leading divergence of the entropy follows an area law and the logarithmic one has a universal coefficient. These divergences are moreover independent of the field strength, so it is meaningful to look at the change in their finite part as we switch on the field. The result is shown in Fig. \ref{fig:thermo} for the charged scalar and in Fig. \ref{fig:thermo2} for the charged spinor.

In the spinor case, both $U$ and $S$ increase monotonically with the background field. For the scalar, this is no longer true when the mass is small, even at small values of the field, where the vacuum instability is still suppressed. The bound is not precisely at $4m^2=1$ as one might have expected. Since a real field in the static patch corresponds to imaginary flux on the sphere, the partition function is in fact complex. We have plotted here only the real parts of the energy and entropy. The interpretation is less clear\footnote{In light of this, it is perhaps relevant to note that there are still differences with a proper thermal state, in which there would be a nonzero Debye mass. As shown in \cite{Popov:2017xut}, the $dS$ isometries instead lead to a vanishing mass.} to us than that of the imaginary part of the partition function, which we can understand in terms of Schwinger pair production. This will be discussed in detail in Sec. \ref{sec: schwinger}, where we will also note that the small mass range is in fact excluded by the FL bound \cite{Montero:2021otb}. 
%%%%%%%%%%%%%%%%%%%%%%%%%%%%%%%%%%%%%
%%%%%%%%%%%%%%%%%%%%%%%%%%%%%%%%%%%%%
\section{\texorpdfstring{$dS$}{} algebra, quasinormal modes, and propagators}\label{sec: algebra}
%%%%%%%%%%%%%%%%%%%%%%%%%%%%%%%%%%%%%
%%%%%%%%%%%%%%%%%%%%%%%%%%%%%%%%%%%%%
In this section, we will write down the symmetry generators which commute with the equations of motion, and obtain the character directly from their action on boundary fields.
The analysis for the case without background flux was performed in \cite{Anninos:2020hfj}. There, it was also shown that the character admits a quasinormal mode expansion, and a quasicanonical partition function picture was put forth. Here we demonstrate that the same expansion holds when including a uniform background field. At the end of this section we also comment on Green functions and find the retarded propagator. Without flux, the conformal equivalence $dS_2 \times S^1 \sim AdS_3$ implies an $SL(2,\mathbb{R})\times SL(2,\mathbb{R})$ symmetry which factorizes the retarded propagator \cite{Anninos:2011af}. The presence of flux on $dS_2$ breaks this symmetry and the structure of the retarded propagator in this case is more involved.  We restrict the discussion to the case of the charged scalar.

%%%%%%%%%%%%%%%%%%%%%%%%%%%%%%%%%%%%%
\subsection{The character from boundary generators}\label{sec: bound}
%%%%%%%%%%%%%%%%%%%%%%%%%%%%%%%%%%%%%
Following App.~A of \cite{Anninos:2020hfj}, we will construct the scalar character $\chi$ by explicitly evaluating the single-particle trace $\tr e^{-iHt}$ in global $dS_2$. The metric is obtained from the one on the sphere by
\begin{equation}
  ds^2 = d\theta^2 + \sin^2 \theta d \varphi^2 \xrightarrow[]{\;\theta = - it +\pi/2\;}   ds^2 = -dt^2 +  \cosh^2(t) d\varphi^2 \,.
\end{equation}
After a further coordinate transformation, 
\begin{equation}
    ds^2 = -dt^2 +  \cosh^2(t) d\varphi^2 \xrightarrow[]{\sech t=\cos\vartheta} ds^2 = (\cos \vartheta)^{-2} (d\vartheta^2 + d\varphi^2) \,,
\end{equation}
so the $\varphi$-circle can be thought of as the future conformal boundary of global $dS_2$. The field strength is proportional to the volume form and we can choose a gauge as follows 
\begin{equation}\label{globalA}
    F = E \cosh t \ dt \wedge d\varphi \,,\quad  A = E \sinh t \ d \varphi \,.
\end{equation}
In solving the Klein-Gordon equation by separating the time dependence of the scalar field 
\begin{equation}\label{eq:klein-gordon}
   \Big((\nabla_\mu - iA_\mu)^2 - m^2 \Big) \phi = 0 \;, \quad \phi(t, \varphi) = y(t) Y(\varphi) \,,
\end{equation}
one finds that the late time behavior is given by
\begin{equation} \label{eq:Delta}
    y(t) \rightarrow c_1 e^{- \Delta t} + c_2 e^{-\bar{\Delta}t} \,, \quad \Delta = \frac{1}{2} + i \eta_\phi \,, \quad \eta_\phi \equiv \sqrt{m^2 + E^2 - \frac{1}{4}} \,,
\end{equation}
where $\bar{\Delta} \equiv 1- \Delta$ and $\Delta$ becomes the scaling dimension of the corresponding boundary state.

\subsubsection*{Boundary states and conformal generators}

The next step is to consider primary states living on the conformal boundary produced by a boundary conformal field $\mathcal{O}(\varphi)$ of dimension $\Delta$ acting on an $SO(1,2)$-invariant global vacuum state $\ket{\text{vac}}$:
\begin{equation}
    \ket{\varphi} \equiv \mathcal{O}(\varphi) \ket{\text{vac}}\,, \quad \braket{\varphi}{\varphi'} = \delta(\varphi - \varphi') \,.
\end{equation}
Before constructing the conformal generators, let us first construct the bulk generators which satisfy the $\mathfrak{so}(1,2)$ algebra
\begin{equation} \label{so21algebra}
    [L_{AB},L_{CD}] = \eta_{BC}L_{AD} +\eta_{DB}L_{CA} + \eta_{AD}L_{BC} + \eta_{CA}L_{DB}\,.
\end{equation}
Being associated to a symmetry, they must commute with the equations of motion \eqref{eq:klein-gordon}. This leads us to the following symmetry generators in the case with flux:
\begin{equation}\begin{split}\label{eq: operators}
    L_{0,1} &\equiv - i \Big[\sin \varphi \partial_t + \cos \varphi \tanh t \Big( \partial_\varphi + \frac{i E}{ \sinh t}\Big)\Big] \,,  \\
    L_{0,2} & \equiv -i \Big[ \cos \varphi \partial_t - \sin \varphi \tanh t \Big(\partial_\varphi +\frac{i E}{ \sinh t}\Big) \Big]\,, \\
    L_{1,2} &\equiv -i \partial_\varphi \,.
\end{split}\end{equation}
The conformal algebra
\begin{equation}
    [H,P] = iP \,, \quad [H,K] = -iK \,, \quad [K,P] = 2iH 
\end{equation}
can be constructed from the above generators by taking the combinations 
\begin{equation}\begin{split} \label{conformalcombinations}
    H = -L_{0,2}\,, \quad
    P =  L_{1,2} + L_{0,1}\,, \quad
    K =  L_{1,2} - L_{0,1} \,.
\end{split}\end{equation}
Evaluated on states $\ket{\varphi}$ living on the future boundary, the conformal generators behave as
\begin{equation}\begin{split}\label{eq:conformal}
    H \ket{\varphi}&= 
    i \Big(\sin \varphi \partial_\varphi + \Delta \cos \varphi \Big) \ket{\varphi} \,, \\
    P \ket{\varphi}&=  i\Big( (1+ \cos \varphi) \partial_\varphi - \Delta \sin \varphi  \Big) \ket{\varphi} \,, \\
    K \ket{\varphi}&=  i\Big( (1-\cos \varphi) \partial_\varphi + \Delta \sin \varphi \Big) \ket{\varphi}\,,
\end{split}\end{equation}
where $\Delta$ is the scaling dimension \eqref{eq:Delta} of the bulk scalar field $\phi$, as determined from the late time behavior of the global $dS_2$ modes. 

\subsubsection*{Calculation of the character}
We are now able to calculate the character $\chi(t) = \tr e^{-iHt}$ as a trace over boundary states. First we take into account the action of $H$ to write
\begin{equation}\label{eq: char}
    \chi(t) = \int^{2\pi}_0 d\varphi \bra{\varphi}e^{-iHt} \ket{\varphi} = \int^{2\pi}_0 d\varphi \;e^{\Delta t}\;\frac{\braket{\varphi}{2 \arctan(e^t \tan\frac{\varphi}{2})}}{(\cos^2 \frac{\varphi}{2}+e^{2t}\sin^2\frac{\varphi}{2})^\Delta}  \,.
\end{equation}
The integral is thus seen to reduce to a sum over the fixed points of $H$, namely $\varphi=0$ and $\pi$. We thus find the character
\begin{equation}\label{eq: charstep2}\begin{split}
    \chi(t) &= \int^{2\pi}_0 d\varphi \;e^{\Delta t}\;\frac{\delta(\varphi-2 \arctan(e^t \tan\frac{\varphi}{2}))}{(\cos^2 \frac{\varphi}{2}+e^{2t}\sin^2\frac{\varphi}{2})^\Delta}\\
    &= \frac{e^{\Delta t}}{|1-e^{t}|}+\frac{e^{-\Delta t}}{|1-e^{-t}|}= \frac{e^{-\Delta t}+ e^{-\bar{\Delta} t}}{|1-e^{-t}|}\,.
    \end{split}
\end{equation}
This is in agreement with the character defined in Sec. \ref{sec: scal1loop} and the character as obtained from a sum over quasinormal modes in the next section. Note that the shift in the operators \eqref{eq: operators} did not influence the eventual result of the calculation \eqref{eq: char}, besides the change in $\Delta$.

Following  \cite{Anninos:2020hfj}, we could have instead conformally mapped to planar boundary coordinates via $x = \tan \frac{\varphi}{2}$, under which the states transform as primaries $(\frac{1}{2}(1+x^2))^\Delta \ket{x} = \ket{\varphi}$. In this basis, $H, P, K$ take the form of shifted dilatation, translation, and special conformal transformations:
\begin{equation}\begin{split}\label{eq: boundarygenerator}
    H \ket{x} = i \Big(x \partial_x + \Delta \Big)\ket{x} \,,  \quad    P \ket{x} = i  \partial_x \ket{x} \,, \quad
    K \ket{x} = i \Big( x^2 \partial_x + 2 \Delta x  \Big) \ket{x} \,.
\end{split}\end{equation}
The calculation proceeds as before, but one has to be careful to include the fixed points at both $x=0$ and $x=\infty$. 

One could have also started with the Wu-Yang potential on the sphere rather than \eqref{globalA}. After Wick rotation to the static patch, the generators \eqref{eq:scalar_angular} would then have a non-Hermitian piece proportional to $E$, which eventually drops out when calculating the trace.

%%%%%%%%%%%%%%%%%%%%%%%%%%%%%%%%%%%%%
\subsection{The character as a quasinormal mode sum}\label{sec: qnm}
%%%%%%%%%%%%%%%%%%%%%%%%%%%%%%%%%%%%%
As we have seen so far, the Harish-Chandra character contains information about the normalizable eigenfunctions used in the heat kernel, but also about the de Sitter isometries \eqref{eq:conformal}. In fact, these symmetries act on 2 towers of quasinormal modes \cite{Sun:2020sgn} as a discrete sum over which the character can be expanded. Although these solutions are not normalizable in the same sense as the modes used in the heat kernel, both sets of modes contain the same information. In $AdS$, this equivalent description in terms of either a continuous or discrete spectrum was first pointed out by \cite{Keeler:2014hba}. In $dS$, it led to the quasicanonical interpretation of the partition function \eqref{eq:scalarchar} in \cite{Anninos:2020hfj}. 

When we add a uniform electric field in $dS_2$, the story does not change and the character still has a quasinormal mode expansion, essentially because the $dS_2$ isometries are unbroken by the field. Let us work in the static patch. The metric in our coordinates is obtained by rotating the spherical one by $t = -i \varphi$ and $x = \cos \theta$. It then takes the usual form
\begin{equation}
    ds^2 = - (1-x^2)  dt^2 + \frac{d x^2}{1-x^2} \,.
\end{equation}
We consider a minimally coupled charged scalar with background flux in the following gauge:
\begin{equation}
    F = E\;  d t \wedge d x \,, \quad A = - E x\ dt \,,
\end{equation}
so that the electric field points in the negative $x$-direction. 

\subsubsection*{Solution of the Klein-Gordon equation with flux}
The solution to the Klein-Gordon equation in this background is well-known \cite{Anninos:2010gh}. Assuming the scalar field $\phi$ has time dependence $\exp(-i \omega t)$, separation of variables $\phi(x,t) = f(x)e^{-i \omega t}$ brings the equations of motion in the form 
\begin{equation}\label{eq: dseom}
    \Big(\frac{(\omega - Ex)^2}{1-x^2}- m^2\Big) f -2x f'  + (1-x^2) f'' = 0 \,,
\end{equation}
which can be further reduced to a hypergeometric equation by substituting
\begin{equation}\label{eq: phisplit}
    f(x) = (1-x)^\alpha (1+x)^\beta g(x)\,, 
    \quad \alpha = -\frac{i}{2} (\omega-E)\,, \quad \beta = -\frac{i}{2} ( E+\omega)\,.
\end{equation}
One solution is then found by taking
\begin{equation}\label{eq: gsol}
   g(x) = {_2F_1}\Big(\frac12 - i {\eta}_\phi - i \omega, \frac12 + i {\eta}_\phi - i \omega; 1-i(\omega-E); \frac{1-x}{2}\Big) \,.
\end{equation}
 Since \eqref{eq: dseom} is real, a second solution for $f(x)$ is simply the complex conjugate of the first one. Looking at the mode \eqref{eq: phisplit} and expanding it around $x=1$, one can see it is purely incoming ($\omega < E$) or outgoing ($\omega > E$) since for $\epsilon \ll 1$
\begin{equation}
   \phi(1-\epsilon, t) \sim \epsilon^{-\frac12 i (\omega-E)}e^{-i\omega t} \sim e^{-i \omega t + i (\omega-E) u} \,,
\end{equation}
with the last term given in coordinates $x = \tanh{u}$. For the complex conjugate mode, the same holds true near $x=-1$. The $u$-coordinates are especially useful since the modes appear as plane waves near the horizons at $u = \pm \infty$. In fact, in terms of these coordinates, \eqref{eq: dseom} at $E=0$ can be recognized as a Schr\"odinger problem with P\"oschl-Teller potential. For non-zero flux, one has a Rosen-Morse potential 
\begin{equation}\label{eq: rosen}
    V(u) = \frac{m^2 + E^2}{\cosh^2 u} + 2E \omega \tanh u \,.
\end{equation}
This type of problem is indeed exactly solvable. In particular for the upside down potential, it features a supersymmetric structure \cite{Cooper:1994eh, kleinert}.

\subsubsection*{Behavior of the normal modes}

\begin{figure}[ht]
    \centering
   \begin{subfigure}{0.4\textwidth}
            \centering
            \includegraphics[width=\textwidth]{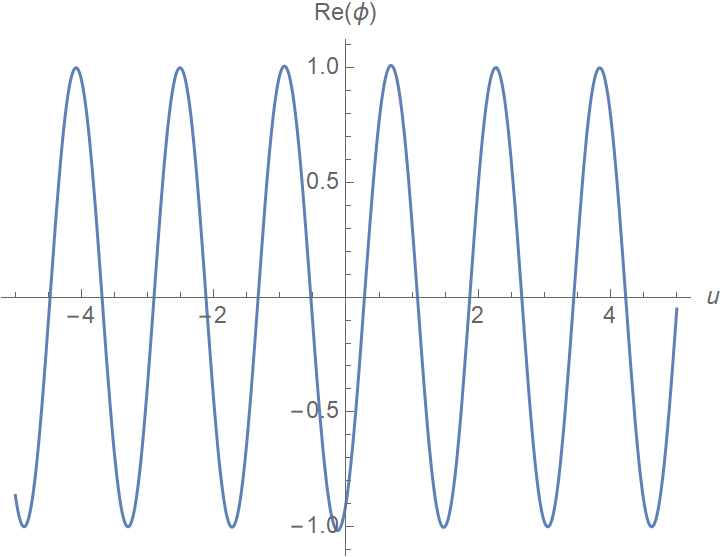}
            \caption[]%
            {{\small mass $m=1$ and frequency $\omega= 4$}}    
            \label{fig:nf1}
        \end{subfigure}
        \hspace{0.5cm}
        \begin{subfigure}{0.4\textwidth}  
            \centering 
            \includegraphics[width=\textwidth]{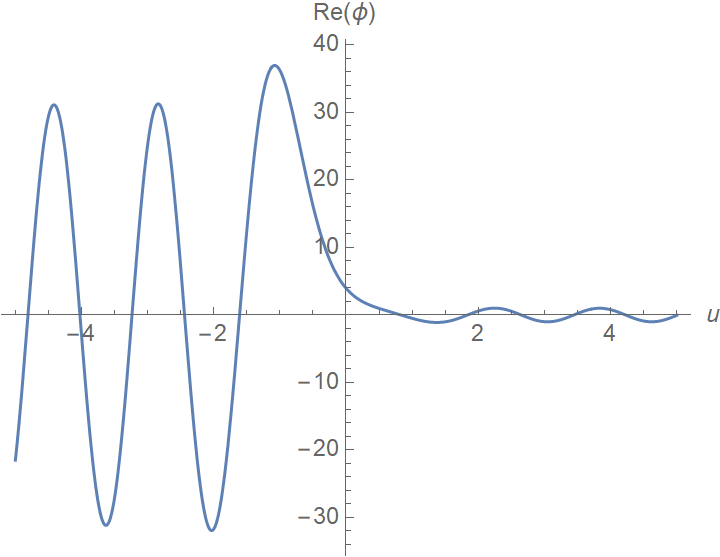}
            \caption[]%
            {{\small mass $m=5$ and frequency $\omega= 4$}}    
            \label{fig:nf2}
        \end{subfigure}
    \caption{We plot the real part of the scalar solution $\phi$ which is outgoing at $x=1$, in the case where the field $E=0$, at time $t=0$. For larger mass transmission is suppressed.}\label{fig:noflux}
\end{figure}

We plot the real part of the fields in Fig. \ref{fig:noflux} in the case without flux, for which the modes become Legendre functions, as in \cite{Akhmedov:2020qxd}. Transmission through the P\"oschl-Teller potential barrier at the origin is less likely for larger mass.

\begin{figure}[ht]
    \centering
   \begin{subfigure}{0.3\textwidth}
            \centering
            \includegraphics[width=\textwidth]{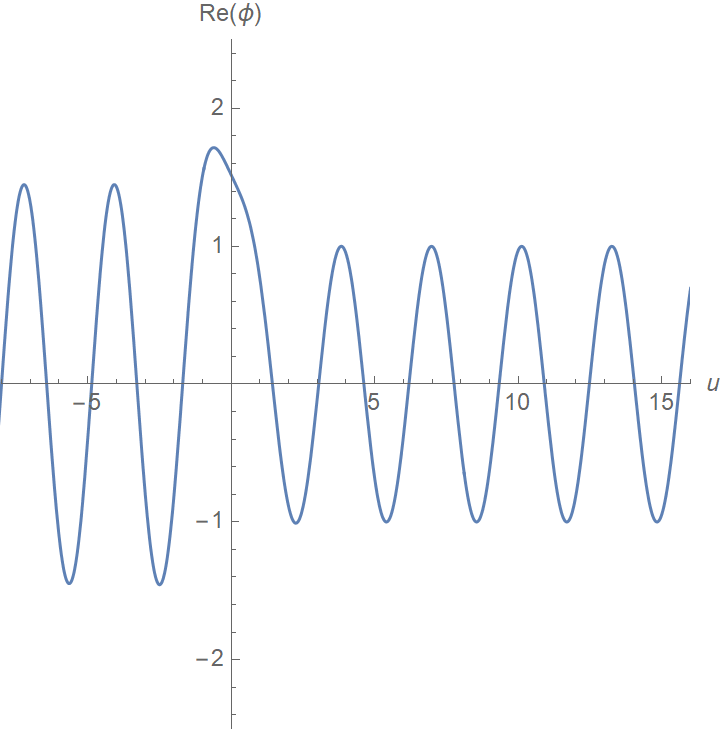}
            \caption[]%
            {{\small $E=0$ }}    
            \label{fig:f1}
        \end{subfigure}
                \hspace{0.1cm}
        \begin{subfigure}{0.3\textwidth}  
            \centering 
            \includegraphics[width=\textwidth]{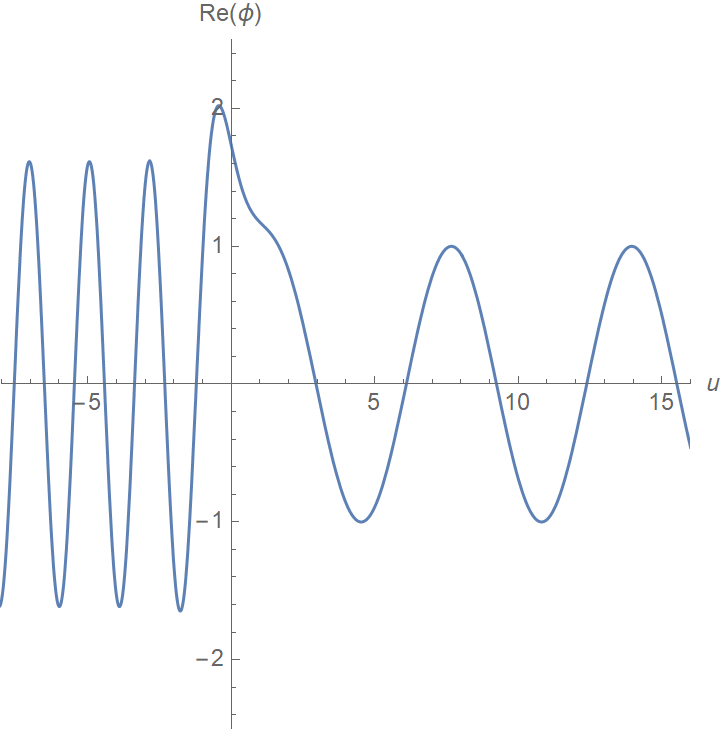}
            \caption[]%
            {{\small $E= 1$}}    
            \label{fig:f2}
        \end{subfigure}
                \hspace{0.1cm}
        \begin{subfigure}{0.3\textwidth}  
            \centering 
            \includegraphics[width=\textwidth]{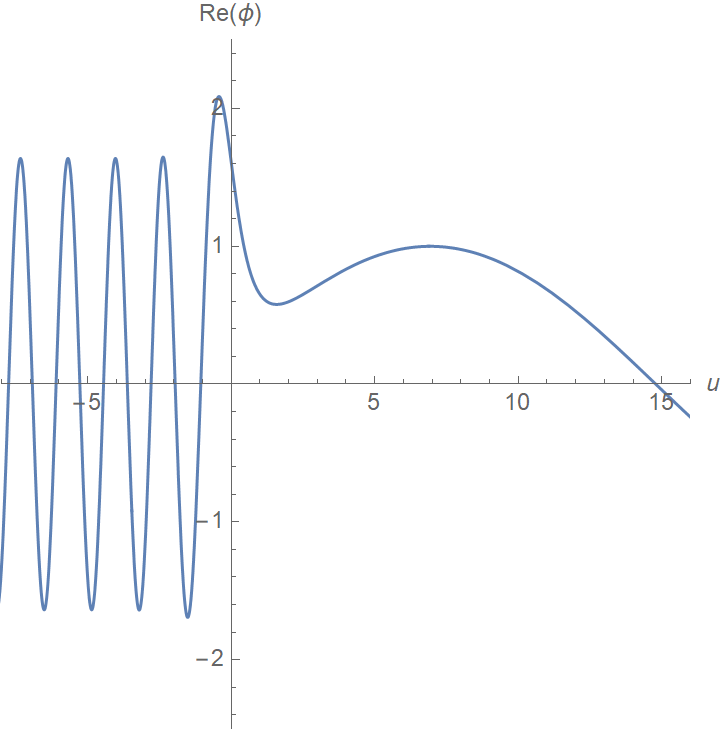}
            \caption[]%
            {{\small $E= 1.8$}}    
            \label{fig:f3}
        \end{subfigure}
    \caption{We plot the real part of the charged scalar $\phi$ which is outgoing at $u=\infty$.  The frequency $\omega = 2$ and mass $m= 2$ are kept fixed, while varying background field $E$. Increasing $E$ reduces the transmission coefficient \eqref{eq: trans}. It increases the momentum of the wave near $x=-1$ and decreases it near $x=1$. At $\omega  = E$, the wave near $x=1$ changes direction.}\label{fig:flux}
\end{figure}

Some solutions with background flux are also plotted in Fig. \ref{fig:flux}. At $\omega = E$, the waves near $u=\infty$ change direction, and the same happens for the waves near $u=-\infty$ when $\omega= -E$. The intermediate $\omega$ interval is known as the level crossing region. Modes in this range appear as having positive or negative energy depending on which side of the potential they are on, and are the ones relevant for superradiance and Schwinger pair production \cite{Damour:1975pr, manogue}. This holds for general step-like potentials, of which the Rosen-Morse potential \eqref{eq: rosen} is one concrete example. From the relative amplitude of the right-moving waves near $u=-\infty$ and $u=\infty$, one finds the transmission coefficient
\begin{equation}\label{eq: trans}
    |T|^2 = \frac{\sinh(\pi(\omega-E) )\sinh(\pi(\omega+E))}{\cosh(\pi(\omega-\eta_\phi))\cosh(\pi(\omega+\eta_\phi))} \,,
\end{equation}
as previously obtained in for instance \cite{Belgiorno:2009pq}. Transmission at fixed $\omega$ decreases when increasing the flux or the mass. Upon reaching the level crossing region, \eqref{eq: trans} becomes negative and one has superradiance. An equivalent system arises in the study of a charged scalar in a rotating Nariai spacetime \cite{Anninos:2010gh}.

\subsubsection*{Quasinormal modes and the character}

The quasinormal modes are now found by demanding that the outgoing solution defined by \eqref{eq: gsol} is also purely outgoing at $x=-1$. Expanding \eqref{eq: phisplit} near $x=-1$, or equivalently $u=-\infty$, we find the following incoming part:
\begin{equation}
    \phi(-1+\epsilon,t)|_\text{incoming} \sim \frac{\Gamma(iE-i\omega)\Gamma(i E +i \omega)}{\Gamma(\bar{\Delta} - i \omega)\Gamma(\Delta - i \omega)}e^{-i \omega t + i (\omega-E) u} \,.
\end{equation}
The above incoming part is absent when we hit the poles of the $\Gamma$-functions in the denominator, thus determining the quasinormal frequencies as
\begin{equation}\label{eq: qnmfreq}
    i\omega_+ = \Delta + n \,, \quad i\omega_{-}  = \bar{\Delta} + n\,, \quad n \in \mathbb{N}\,.
\end{equation}
Summing over these modes, we indeed retrieve the previously obtained scalar character
\begin{equation}
   \chi(t) = \sum^\infty_{n = 0} e^{-i\omega_+ t} + e^{-i\omega_{-} t} = \frac{e^{-\Delta t} + e^{-\bar{\Delta} t}}{1-e^{-t}}  \,,
\end{equation}
as in \eqref{eq:scalarchar} and \eqref{eq: charstep2}. 

%%%%%%%%%%%%%%%%%%%%%%%%%%%%%%%%%%%%%
\subsection{Retarded propagator and Green functions}
%%%%%%%%%%%%%%%%%%%%%%%%%%%%%%%%%%%%%
The results of the previous section would give the impression that all of the algebraic structure present without flux survives when we turn on a background electric field. This is not the case, however. In this section, we will calculate the retarded static patch propagator following \cite{ Anninos:2011af, Anninos:2017hhn}. We will see that, unlike in the case without flux, it no longer factorizes in an obvious way. For completeness, we will also derive the coordinate space Green function, the coincidence limit of which will be used in App. \ref{app: B} to verify the results of Sec. \ref{sec: schwingscal}.

\subsubsection*{Retarded propagator on the static patch worldline}
\begin{figure}[ht]
     \centering
   \begin{subfigure}{0.3\textwidth}
            \centering
            \includegraphics[width=\textwidth]{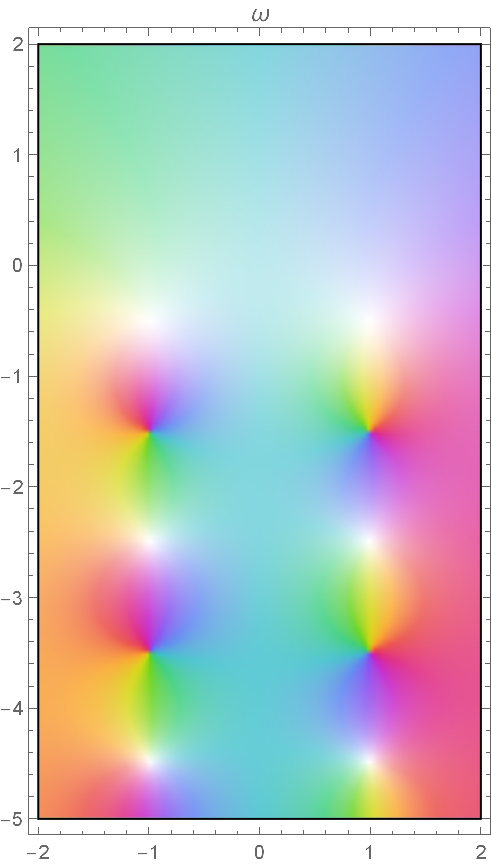}
            \caption[]%
            {{\small  $E=0$}}    
        \end{subfigure}
        \hspace{0.5cm}
        \begin{subfigure}{0.3\textwidth}  
            \centering 
            \includegraphics[width=\textwidth]{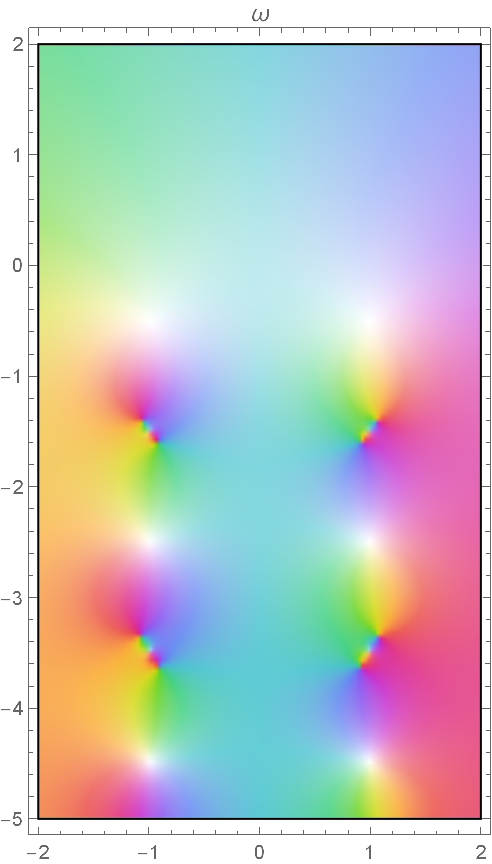}
            \caption[]%
            {{\small $E= 0.1$}}    
        \end{subfigure}
                \hspace{0.5cm}
        \begin{subfigure}{0.3\textwidth}  
            \centering 
            \includegraphics[width=\textwidth]{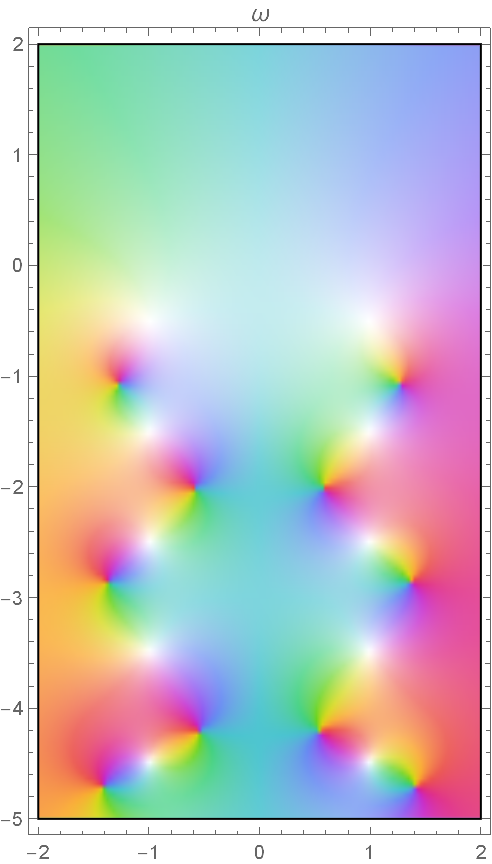}
            \caption[]%
            {{\small $E= 0.5$}}    
        \end{subfigure}
     \caption{A phase plot of the retarded static patch propagator $G_R$ at coincidence, see \eqref{eq: retarded}, as a function of the frequency $\omega$, evaluated at $\Delta =\frac12 + i$, for different values of the electric field $E$. All poles are located on the lower half of the complex plane. The 2 towers of quasinormal modes \eqref{eq: qnmfreq} are clearly visible. For $E=0$, one alternatingly has poles and zeros at these frequencies. When switching on the field, the zeros split up, and all quasinormal frequencies appear as poles of $G_R$.}
     \label{fig: complexplot}
 \end{figure}

As noted in the previous section, the coordinate transformation $x= \tanh u$ brings us to a Schr\"odinger type problem with Rosen-Morse potential. Let us write down the modes which are going into the left and right horizon respectively:
\begin{equation}
    \phi_{l,r}(u,t) = e^{-i\omega t-i E u}(\cosh u)^{i\omega}g_{l,r}(u)\;,
\end{equation}
where
\begin{equation}\begin{split}
    g_l(u) &= {}_{2}F_1\Big(\bar{\Delta}-i\omega,\Delta+i\omega;\;1-i(\omega+E);\;\frac{e^{u}}{2} \sech u\Big) \,,\\
    g_r(u) &= {}_{2}F_1\Big(\bar{\Delta}-i\omega,\Delta+i\omega;\;1-i(\omega-E);\;\frac{e^{-u}}{2}\sech u\Big)\,,
\end{split}
\end{equation}
 which are related by flipping the sign of $u$ and $E$.  
Using the Wronskian method \cite{MathematicsforthePhysicalSciences}, we can then write down the retarded propagator for $-\infty < u < u' < \infty$:
\begin{equation}\label{eq: retardedwronsk}
    G_R (u,u',\omega) = \frac{\phi_l(u)\phi_r(u')}{W(\varphi_l,\varphi_r)} \,.
\end{equation}
The Wronskian $W(f,g)= fg'-gf'$ is independent of $u$; in particular we can evaluate it near the origin. Using the expansion
\begin{equation}
     \phi_l(u) = \beta + \alpha u + O(u^2)\,, \quad \phi_r(u) = \beta' + \alpha' u + O(u^2)\,,
\end{equation}
we find the coincidence limit of the retarded propagator
\begin{equation}\label{eq: retarded}
    \lim_{u\to0}G_R(u,0,\omega) = \frac{\beta \beta'}{\beta \alpha'-\alpha \beta'}\,.
\end{equation}
The expressions for $\alpha$ and $\beta$ are\footnote{Here we should note that in higher dimensions, one instead imposes outgoing boundary conditions at the horizon, which is now connected,  and normalizability at the origin. Extending this terminology to $d=1$, we would have $\beta (\alpha)$ correspond to the (non)-normalizable mode. Constructing the retarded propagator based on the outgoing mode at the horizon and normalizable one at the origin then leads to $G_R = \beta/\alpha$, reminiscent of the holographic vev/source presciption \cite{Anninos:2011af}. However, in the case of $d=1$, these boundary conditions are not adequate because of the presence of both a left and right horizon \cite{Anninos:2017hhn}.}
\begin{equation}\label{eq: alphabeta}
    \begin{split}
        \beta  = &{}_{2}F_1\Big(\bar{\Delta}-i\omega,\Delta+i\omega;\;1-i(\omega+E);\;\frac12 \Big)\,,\\
        \alpha  =\frac{(\bar{\Delta}-i\omega)(\Delta-i\omega)}{2(1-i(\omega+E))}& {}_{2}F_1\Big(1+\bar{\Delta}-i\omega,1+\Delta+i\omega;\;2-i(\omega+E);\;\frac12 \Big) - i \beta E \,,
    \end{split}            
\end{equation}
 and $\alpha',\;\beta'$ are related to $\alpha,\;\beta$ by flipping the signs of $E$ and $\alpha$. The analytic structure of the retarded propagator is shown in Fig. \ref{fig: complexplot} for different values of the electric field, and its poles correspond to quasinormal frequencies. With non-vanishing flux, all of them are present, whereas at $E=0$, \eqref{eq: retarded} simplifies to \cite{Anninos:2011af, Anninos:2017hhn}
\begin{equation}\label{eq:retardednoflux}
 G_R(0,0,\omega) = \frac{\Gamma(\frac12(\Delta-i\omega))\Gamma(\frac12(\bar{\Delta}-i\omega))}{4\Gamma(\frac12(1+\Delta-i\omega))\Gamma(\frac12(1+\bar{\Delta}-i\omega))} \,,
 \end{equation}
in which case $G_R$ has poles only at half of the quasinormal frequencies, namely $\omega \in -i(\Delta + 2\mathbb{N})$. The reason for the factorization $G_R = P(\Delta,\omega)P(\bar{\Delta},\omega)$ in \eqref{eq:retardednoflux} is the existence of an $SL(2,\mathbb{R})\times SL(2,\mathbb{R})$ symmetry, due to the fact that $dS_2\times S^1$ is conformal to $AdS_3$ \cite{Anninos:2011af}. The presence of flux on $dS_2$ breaks this symmetry and the general expression \eqref{eq: retarded} does not seem to have this nice structure\footnote{Similarly, the supersymmetric structure \cite{Anninos:2011af} found for the conformal scalar $2i\nu = 1$ rests on the fact that the 2 towers of quasinormal modes coincide. In general, requiring $2i\eta = 1$ would impose either $m=0, E=0$ as before, imaginary flux in $dS$, or negative mass squared.}.
 
\subsubsection*{Coordinate space Green Function}
For generality and to double-check our results in Sec. \ref{sec: schwingscal} using the formal relation 
\begin{equation}\label{eq: logZvsGF}
    \frac{\partial}{\partial m^2} \log Z = -\int d^2x \sqrt{g} \  G(x,x) \,,
\end{equation}
we will obtain in this section the coordinate space Green function at coincidence. The real part of \eqref{eq: logZvsGF} is divergent, but its imaginary part is finite and should satisfy this consistency relation, as we verify in App. \ref{app: B}. The calculation for pure de Sitter can be found in \cite{Candelas:1975du}. 

First, let us determine the Green function on the sphere, which satisfies
\begin{equation} \label{Wightmaneq}
    \Big[ (\vec{\nabla}-i \vec{A})^2-m^2\Big]G(X,Y)=\frac{1}{\sqrt{g}}\delta^{(2)}(X-Y) \,.
\end{equation}
Considering $X,Y$ as unit vectors in $\mathbb{R}^3$, one constructs the invariant distance 
\begin{equation}
    P \equiv \delta_{AB}X^AY^B \,,
\end{equation}
which is the cosine of the angle between $X$ and $Y$. By symmetry, the Green function depends only on $P$, and away from coincidence, \eqref{Wightmaneq} can be written as 
\begin{equation}
    \Big[(1-P^2)\partial_P^2 -2P \partial_P - B^2\frac{1-P}{1+P} -m^2\Big]G(P)=0\,.
\end{equation}
We are interested in the solution which is regular when $X$ and $Y$ are antipodal, corresponding to $P=-1$. Defining $z \equiv \frac{1+P}{2}$, this solution is given by
\begin{equation}\begin{split}
    G(z) =
    C\, z^{B} \ {}_2F_1\Big(\Delta+B,\bar{\Delta}+B,1+2B,z\Big) \,.
\end{split}\end{equation}
To fix the prefactor $C$, we need to reproduce the $\delta$-function in \eqref{Wightmaneq}. This is done by mapping the logarithmically divergent short-distance behavior of $G$ to that in flat space. Using (15.3.10) of \cite{abramowitz+stegun}, in the $z\rightarrow1$ limit, we find
\begin{equation}
    G(z \rightarrow 1) \approx -C\; \frac{\Gamma(1+2B)}{\Gamma(\Delta + B)\Gamma(\bar{\Delta} + B)}\Big(2 \gamma + \psi(\Delta + B) + \psi(\bar{\Delta} + B) + \log(1-z)  \Big) \,.
\end{equation}
In 2-dimensional flat-space, the coincident limit of the scalar Green function is \cite{DiFrancesco:1997nk}
\begin{equation}
    G_\text{flat}(z) \approx -\frac{1}{4\pi} \Big(2 \gamma + \log m^2 + \log(1-z) \Big)\,.
\end{equation}
We should therefore choose 
\begin{equation}
    C= \frac{1}{4 \pi} \frac{\Gamma(\Delta + B)\Gamma(\bar{\Delta} + B)}{\Gamma(1+2B)}\,,
\end{equation}
so that
\begin{equation}\label{Wightman_solved}
    G(z) = \frac{1}{4\pi}\frac{\Gamma(\Delta + B)\Gamma(\bar{\Delta} + B)}{\Gamma(1+2B)}z^{B} {}_2F_1\Big(\Delta+B,\bar{\Delta}+B,1+2B,z\Big)\,.
\end{equation}
We can now analytically continue from $S^2$ to $dS_2$ by rotating $B\rightarrow i E$ and replacing the invariant distance on $S^2$ by the invariant distance $P=\eta_{AB}X^{A}Y^{B}$ in $dS_2$. In static patch coordinates $x=\cos\theta$, it is given by
\begin{equation}
    P = \sin\theta_1 \sin\theta_2\cosh{(t_1-t_2)} +\cos\theta_1 \cos\theta_2 \,.
\end{equation}
 In App. \ref{app: B}, we verify that \eqref{Wightman_solved} satisfies \eqref{eq: logZvsGF}, providing an alternative derivation of the results obtained by heat kernel regularization in Sec. \ref{sec: schwingscal}.

%%%%%%%%%%%%%%%%%%%%%%%%%%%%%%%%%%%%%
%%%%%%%%%%%%%%%%%%%%%%%%%%%%%%%%%%%%%
\section{Schwinger pair production}\label{sec: schwinger}
%%%%%%%%%%%%%%%%%%%%%%%%%%%%%%%%%%%%%
%%%%%%%%%%%%%%%%%%%%%%%%%%%%%%%%%%%%%
In the presence of a background electric field, the vacuum is subject to Schwinger pair production. This phenomenon is captured by the imaginary part of the partition function \cite{Schwinger:1951nm, Damour:1975pr, manogue}. Here, making use of the character integrals from Sec. \ref{sec: char}, we evaluate the vacuum persistence for the $dS_2$ static patch by analytic continuation from $S^2$. A similar method was applied for $AdS_2$ and $\mathbb{H}^2$ in \cite{Pioline:2005pf}. We include the effect of the static patch thermal background by retaining periodicity in Euclidean time, and make a comparison with the non-thermal results of \cite{Belgiorno:2009pq,Belgiorno:2009da}. Our setup is more closely related to the work of \cite{Frob:2014zka}, who found the current created in the $dS_2$ planar patch in the global vacuum. In particular, the semiclassical contributions due the screening and antiscreening instantons \cite{Garriga:1993fh, Garriga:1994bm} appear with the same prefactor and relative sign in our result for the vacuum persistence as in the current found in \cite{Frob:2014zka}. This is not the case for the non-thermal static patch results \cite{Belgiorno:2009pq,Belgiorno:2009da}. 

An interesting physical setup in which charged particles in a uniform electric field on $dS_2$ appear comes from dimensionally reducing charged or rotating Nariai spacetimes \cite{Belgiorno:2009pq, Belgiorno:2009da, Anninos:2010gh}. We note in particular that the scalar mass region in which there is IR hyperconductivity \cite{Frob:2014zka} (see Fig. \ref{fig:imbehaviorscal}) is excluded by the recently proposed FL bound \cite{Montero:2019ekk, Montero:2021otb}. We will also apply the character formalism as developed for $AdS$ in \cite{Sun:2020ame} to understand pair creation around the $AdS_2$ black hole and clarify a result from \cite{Pioline:2005pf}.  In $AdS$, there is a threshold for pair creation, unlike in $dS_2$ where it happens for any non-zero value of the electric field.

%%%%%%%%%%%%%%%%%%%%%%%%%%%%%%%%%%%%%
\subsection{Vacuum persistence for charged scalars}\label{sec: schwingscal}
%%%%%%%%%%%%%%%%%%%%%%%%%%%%%%%%%%%%%
For the scalar partition function, upon rotating $B \to i E$, corresponding to a real electric field in $dS_2$, we find that the imaginary part of the 1-loop partition function \eqref{eq:scalar_fin} is finite. We can safely take $\epsilon \to 0$, resulting in
\begin{equation}\label{eq:impart}
     \Im[\log Z_\phi] =\int^\infty_{-\infty} \frac{d t}{2t}\frac{\cos{2\eta_\phi t}}{\sinh{t}}(2E\cos{2 E t} - \coth{t}\sin{2E t}) \,.
\end{equation}
Note that we have extended the integral over the entire real axis, using the fact that the integrand is even and regular at zero. The poles are located at $t \in i\pi \mathbb{Z}\setminus\{0\} $. Summing over the residues, we find
\begin{equation}\begin{split}\label{eq:imscalar}
    2\pi \Im[\log Z_\phi] =\; & \text{Li}_2(-e^{2\pi(E-\eta_\phi)})-\text{Li}_2(-e^{-2\pi(E+\eta_\phi)})\\ &+ 2\pi \eta_\phi\;\text{Li}_1(-e^{2\pi(E-\eta_\phi)})- 2\pi \eta_\phi\;\text{Li}_1(-e^{-2\pi(E+\eta_\phi)})\,,\end{split}
\end{equation}
which is odd in $E$. In particular, this means the imaginary part vanishes when the flux does. Note also that the arguments of the polylogs in \eqref{eq:imscalar} are negative, meaning that the different semiclassical instanton contributions carry an alternating sign, as in flat space. Alternatively, one can start from the exact expression \eqref{eq:scalar_fin}, replace $B$ by $iE$, and use the relation 
\begin{equation}\label{eq:hurpoly}
    \text{Li}_s(z) + (-1)^s \text{Li}_s(z^{-1}) = \frac{(2 \pi i)^s}{\Gamma(s)}\zeta\Big(1-s, \frac12 + \frac{\log(-z)}{2\pi i}\Big)
\end{equation}
between Hurwitz $\zeta$-functions and polylogarithms to arrive at the same result \eqref{eq:imscalar}. The semiclassical interpretation of the $\text{Li}_1$ terms will be discussed in Sec. \ref{sec:compnonthermal} . The $\text{Li}_2$ terms also arise as surface corrections in Rindler space \cite{Gabriel:1999yz} and $AdS_2$ \cite{Pioline:2005pf}.

The correct flat space limit is obtained by first reinstating the de Sitter length $l$ by $m\to ml$ and $E\to El^2$ and then taking $l \to \infty$, such that
\begin{equation}
    \eta_\phi = \sqrt{m^2l^2 + E^2 l^4 - \frac14} \approx E l^2 + \frac{m^2}{2E} \,.
\end{equation}
Taking this limit in \eqref{eq:imscalar}, the polylogs with $E+\tilde{\eta}_\phi$ in the exponent vanish and the $\text{Li}_1$ contribution dominates in the remaining two terms because of the prefactor $\eta_\phi$. Expanding this dominating contribution, we find
\begin{equation}\label{eq:flatscalar}
    \Im[\log Z_\phi] \to E l^2 \sum^\infty_{k=1} \frac{(-1)^k}{k}e^{-\pi k m^2/E} \,.
\end{equation}
Identifying the Euclidean volume $V=4\pi l^2$,  this is seen to equal the flat instanton sum \cite{Schwinger:1951nm} for Schwinger pair production
\begin{equation}\label{eq:flatim}
    \Im[\log Z_{\phi, \text{flat}}] = \frac{EV}{4\pi} \sum^\infty_{k=1} \frac{(-1)^k}{k}e^{-\pi k m^2/E},
\end{equation}
 as obtained by summing the residues of \eqref{eq:flatscal}.

%%%%%%%%%%%%%%%%%%%%%%%%%%%%%%%%%%%%%
\subsection{Vacuum persistence for charged spinors}
%%%%%%%%%%%%%%%%%%%%%%%%%%%%%%%%%%%%%
In the spinor case, we can similarly extract the imaginary part of the effective action. In heat kernel regularization, the spinor character formula gave the final result \eqref{eq:spinor_fin} at 1-loop. After rotating $B \to i E$, the imaginary part becomes 
\begin{equation}\begin{split}
   \Im[ \log Z_\psi]=  &-2\Im\Big[\zeta'(-1,1+i(E+\eta_\psi)) - i\eta_\psi\; \zeta'(0,1+i(E+\eta_\psi))\Big] \\
   &-2\Im\Big[\zeta'(-1,1+i(E-\eta_\psi)) + i\eta_\psi\; \zeta'(0,1+i(E-\eta_\psi))\Big]\,,  \end{split} 
\end{equation}
where $\eta_\psi = \sqrt{m^2 + E^2}$, which can again be written in terms of polylogs. Making use of the relation \eqref{eq:hurpoly} and the identity
\begin{equation}
   \text{Li}_2(z) = - \text{Li}_2\Big(\frac{1}{z}\Big) - \frac{\pi^2}{6}-\frac{1}{2} \log^2(- z) \,,
\end{equation}
we find
\begin{equation}\label{eq:imspinor}\begin{split}
    2 \pi \Im[\log Z_\psi] =  \ &\text{Li}_2(e^{-2\pi(E+  \eta_\psi)}) -\text{Li}_2(e^{2\pi(E-  \eta_\psi)})\\
     &+2 \pi \eta_\psi \; \text{Li}_1(e^{-2\pi (E+\eta_\psi)})-2 \pi \eta_\psi \; \text{Li}_1(e^{2\pi(E- \eta_\psi)})\;.  
\end{split}\end{equation}
The instanton contributions for spinor pair creation thus all come with the same sign, as in flat space limit, which is obtained by taking $l \to \infty$ keeping $E,m$ fixed, yielding 
\begin{equation}
    \Im[\log Z_{\psi, \text{flat}}] = -\frac{EV}{4\pi} \sum^\infty_{k=1} \frac{1}{k}e^{-\pi k m^2/E}\,.
\end{equation}

%%%%%%%%%%%%%%%%%%%%%%%%%%%%%%%%%%%%%
\subsection{Discussion and comparison to the non-thermal result} \label{sec:compnonthermal}
%%%%%%%%%%%%%%%%%%%%%%%%%%%%%%%%%%%%%

In this section, our aim is to obtain a better physical understanding of the results  \eqref{eq:imscalar} and \eqref{eq:imspinor}
for the imaginary part of the 1-loop partition function in a background electric field. We shall do so by looking at the flat space and semiclassical limits and by comparing to other results in the literature. Of particular interest in this regard are the results of \cite{Belgiorno:2009pq, Belgiorno:2009da}, which can in fact be written more concisely as
\begin{equation}
      \begin{split}\label{eq:imscalarIT}
   \qquad 2\pi \Im[\log Z'_\phi] =\; &-2\pi E\big[\text{Li}_1(-e^{-2\pi(E+\eta_\phi)})+\text{Li}_1(-e^{2\pi(E-\eta_\phi)}) \big] \\ 
    &+\text{Li}_2(-e^{-2\pi(E-\eta_\phi)}) - \text{Li}_2(-e^{2\pi(E+\eta_\phi)})\,,\end{split} 
\end{equation}
and
\begin{equation}
   \begin{split}\label{eq:imspinorIT}
        2 \pi \Im[\log Z'_\psi] = \ & 2\pi E \big[ \text{Li}_1(e^{-2\pi (E+ \eta_\psi)})+\text{Li}_1(e^{2\pi(E-  \eta_\psi)}) \big] \\
    &+\text{Li}_2(-e^{-2\pi(E+  \eta_\psi)}) -\text{Li}_2(-e^{2\pi(E-  \eta_\psi)})\,,
\end{split} 
\end{equation}
for the scalar and spinor respectively. The derivation in \cite{Belgiorno:2009pq, Belgiorno:2009da} was based on a similar heat kernel calculation after Wick rotation from the static patch, but without making the Euclidean time  $\varphi$ compact. Therefore, \eqref{eq:imscalarIT} and \eqref{eq:imspinorIT} pertain to the non-persistence of the static patch vacuum. Our sphere results \eqref{eq:imscalar} and \eqref{eq:imspinor} on the other hand capture the physics of the global vacuum as seen by a static patch observer.  In our calculation, $\varphi$ was indeed made compact and this periodicity in imaginary time encodes the de Sitter temperature. From the static patch point of view, the global vacuum is contains a thermal background of particles. The relation between the results is in fact similar to how Minkowski vacuum results are obtained by thermalizing Rindler ones \cite{Rajeev:2019bzv}. We will argue this further by comparing to \cite{Frob:2014zka}, who calculated the current observed in the $dS_2$ planar patch in the global vacuum. Let us first make a few other observations.

\begin{figure}[ht]
    \centering
   \begin{subfigure}{0.45\textwidth}
            \centering
            \includegraphics[width=\textwidth]{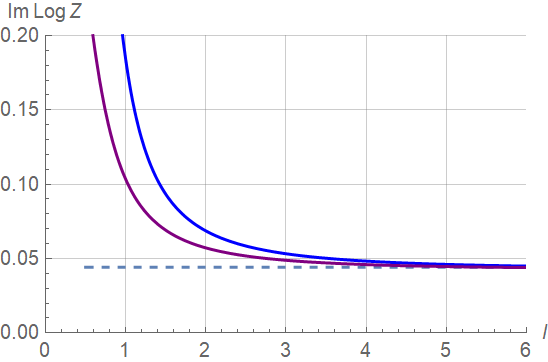}
            \caption[]%
            {{\small  Both \eqref{eq:imscalar} (blue) and \eqref{eq:imscalarIT} (purple) approach the flat space vacuum persistence \eqref{eq:flatim} (dashed) as $l$ is increased, at $E=2$.}}  
        \end{subfigure}
        \hspace{0.5cm}
        \begin{subfigure}{0.45\textwidth}  
            \centering 
            \includegraphics[width=\textwidth]{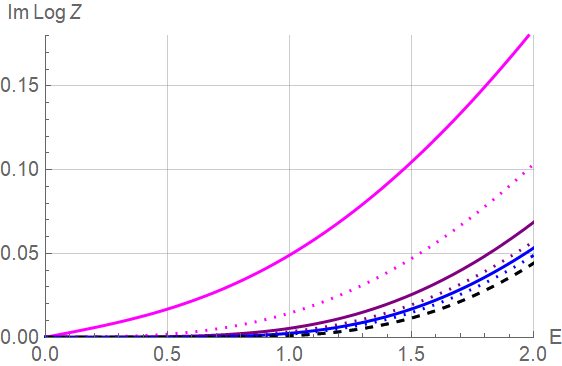}
            \caption[]%
            {{\small $\text{Im}[\log Z]$ as function of  $E$, for $l=1, 2, 3$ (magenta, purple, blue). Solid, dotted and dashed lines are \eqref{eq:imscalar},  \eqref{eq:imscalarIT} and \eqref{eq:flatim}.}}    
        \end{subfigure}
    \caption{Both the thermal \eqref{eq:imscalar} and   non-thermal \eqref{eq:imscalarIT} vacuum persistence have the correct flat space limit \eqref{eq:flatim}. The approach is slower in the thermal case. In both graphs, the mass $m=1$ is kept fixed. $E$ is the electric field and the de Sitter length is denoted by $l$. }\label{fig:flatlimit}
\end{figure}

\begin{figure}[ht]
    \centering
   \begin{subfigure}{0.45\textwidth}
            \centering
            \includegraphics[width=\textwidth]{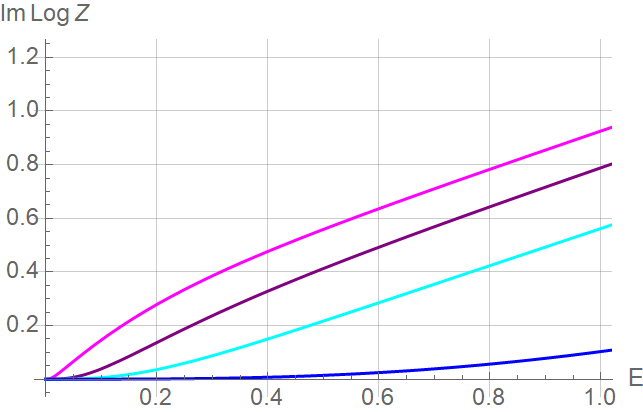}
            \caption[]%
            {{\small non-thermal result \eqref{eq:imscalarIT}}}    
        \end{subfigure}
        \hspace{0.5cm}
        \begin{subfigure}{0.45\textwidth}  
            \centering 
            \includegraphics[width=\textwidth]{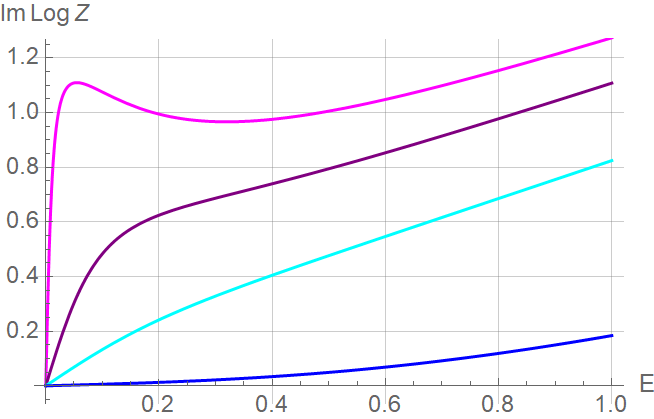}
            \caption[]%
            {{\small thermal result \eqref{eq:imscalar}}}  \label{fig:imbehaviorscalb}  
        \end{subfigure}
    \caption{The vacuum non-persistence $\text{Im}[\log Z]$, as a function of the electric field $E$, at fixed mass $m=0.1,$ $0.3,$ $0.5$ and $1$, from magenta to blue (top to bottom). The non-thermal results are always monotonic. The thermal enhancement is most clear at small field strength and for small masses it leads to the IR hyperconductivity noted in \cite{Frob:2014zka}. }\label{fig:imbehaviorscal}
\end{figure}

\begin{figure}[ht]
    \centering
   \begin{subfigure}{0.45\textwidth}
            \centering
            \includegraphics[width=\textwidth]{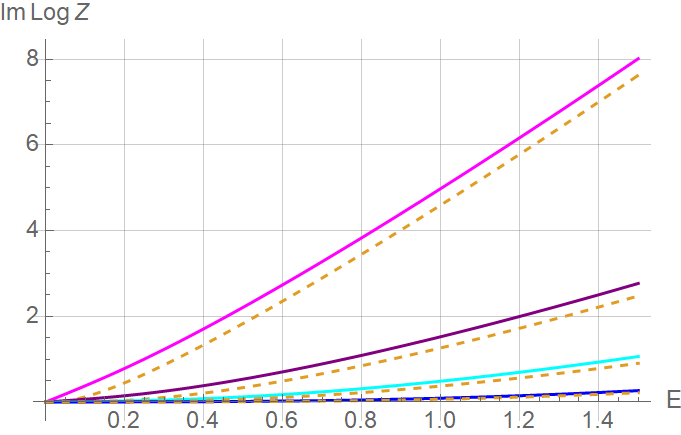}
            \caption[]%
            {{\small More pairs are created as $E$ increases. Ranging from magenta to blue, $m=0.05,$ $0.3,$ $0.6$ and $1.$}}    
        \end{subfigure}
        \hspace{0.5cm}
        \begin{subfigure}{0.45\textwidth}  
            \centering 
            \includegraphics[width=\textwidth]{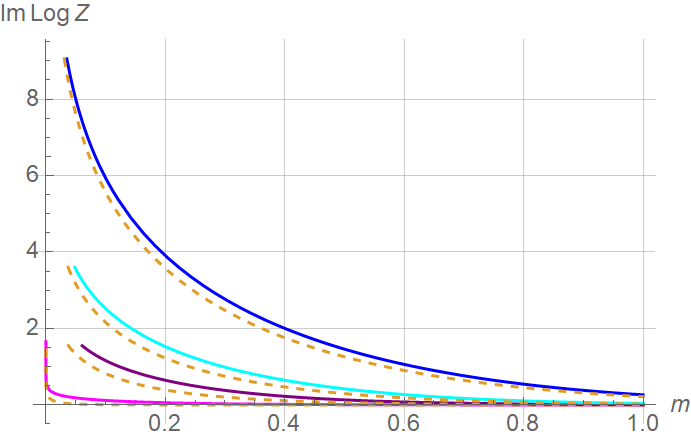}
            \caption[]%
            {{\small Less pairs are created as $m$ increases. Ranging from magenta to blue, $E=0.05,$ $0.4,$ $0.75$ and $1.5.$}}    
        \end{subfigure}
    \caption{The non-persistence of the vacuum in the spinor case, as function of electric field $E$ and mass $m$. The solid lines are \eqref{eq:imspinor}. The dashed ones are \eqref{eq:imspinorIT}. The non-thermal result is always closer to the flat space one. Note that unlike the scalar case, both spinor results are monotonic. Moreover, the imaginary part blows up as the mass goes to zero. For the scalar, this limit is finite.}\label{fig:imbehaviorspin}
\end{figure}

First note that when written in the form \eqref{eq:imscalarIT} and \eqref{eq:imspinorIT}, it is clear that these reduce to the correct flat space limit, just as our results did. This is visualized for the scalar in Fig. \ref{fig:flatlimit}. Further, the expressions for $\text{Im}[\log Z]$ are all odd in $E$. This means that the $\text{Li}_1$ terms come with a different relative sign depending on whether the prefactor is $\eta$, as in \eqref{eq:imscalar} and \eqref{eq:imspinor}, or $E$, as in \eqref{eq:imscalarIT} and \eqref{eq:imspinorIT}, which will be relevant when discussing the semiclassical limit. Another important distinction is that when expanding the polylogs in our spinor result \eqref{eq:imspinor}, all terms come with the same sign, as in \cite{Pioline:2005pf}. In \eqref{eq:imspinorIT} however, they come with alternating sign. In Fig. \ref{fig:imbehaviorscal} and \ref{fig:imbehaviorspin}, one can see the behavior of the rate of Schwinger pair production. The relative sign difference between the leading $\text{Li}_1$ terms and the $\text{Li}_2$ corrections in our scalar result \eqref{eq:imscalar}, compared to \eqref{eq:imscalarIT}, leads to an increased non-persistence of the vacuum at low mass and small fields, where one intuitively expects the effect of the thermal background to be most pronounced. This non-monotonicity seems to be the same phenomenon as the IR hyperconductivity peak found by \cite{Frob:2014zka} in the planar patch setup for $\eta^2 <0$.  In the spinor case, the difference between \eqref{eq:imspinor} and \eqref{eq:imspinorIT} is less noticeable and both are in fact monotonic. In all cases, the thermal result is larger than the non-thermal one, as one would expect.

A transmission coefficient calculation of the non-persistence of the vacuum due to scalar pair creation, along the lines of \cite{manogue}, would start from an integral
\begin{equation}
    W = \frac12\int^{eE}_{-eE} d\omega \log(1+|T|^2)
\end{equation}
 over the level crossing region. Taking $T$ to be the transmission coefficient \eqref{eq: trans}  gives precisely \eqref{eq:imscalarIT} \cite{Belgiorno:2009pq}. A similar calculation was performed for the spinor in \cite{Belgiorno:2009da}. This is equivalent to a Bogoliubov coefficient type calculation in the static patch vacuum, which indeed does not take into account the thermal background. As demonstrated in \cite{Belgiorno:2009pq, Belgiorno:2009da} this is equivalent to taking the Euclidean time non-compact in the $\zeta$-function approach. The non-persistence of the vacuum thus comes from integrating $\propto \log(1+|T|^2)$, and one should keep in mind that the actual pair production rate responsible for the observed current is given instead by the first instanton term $\propto |T|^2$ alone \cite{Damour:1975pr, Cohen:2008wz}. For the spinor, the vacuum persistence is determined by $\log(1- |T|^2)$ instead, which explains why in Fig. \ref{fig:imbehaviorspin}, unlike the scalar case, the result is not finite as $m\to 0$.

Let us now discuss the semiclassical limit. Starting from the worldline action of a charged particle of mass $m$ on the sphere with magnetic field $B$ and solving the geodesic equations after conveniently orienting the axes, one finds a family of solutions 
\begin{equation}\label{eq: solution}
    \theta(s) = \theta_0 \;,\quad
    \varphi(s) = -\frac{B}{m \cos \theta_0} s + \varphi_0 \,.
\end{equation}
The geodesics are circles with radius $\theta_0$. That they always fit on the sphere seems to be a statement without much content, but can be appreciated when noting that in flat space, at high temperature and small field strength, it is possible that the circular geodesics do not fit on the cylinder; instead one finds lens-shaped instantons \cite{Brown_2018} consisting of two less-than-semicircular arcs. Extremizing over $\theta_0$ leads to the action
\begin{equation}
    S_\pm = 2 \pi \Big( \sqrt{m^2+B^2}\pm B \Big) = 2\pi (\eta \pm B)\,.
\end{equation}
for the (anti)screening instanton \cite{Garriga:1993fh, Garriga:1994bm}. These reproduce the leading terms in the series expansion of the $\text{Li}_1$ in the 1-loop $\text{Im}[\log Z]$ as given by \eqref{eq:imscalar} and \eqref{eq:imspinor}:
\begin{equation}\label{eq: semiclass}
\text{Im}[\log Z] \propto \eta (e^{-S_-}-e^{-S_+}) + \text{corrections}\,.
\end{equation}
The instantons come with prefactor $\eta$, as also derived in the semiclassical regime by \cite{Garriga:1993fh, Kim:2015azt}. This is to be contrasted with the non-thermal results \eqref{eq:imscalarIT} and \eqref{eq:imspinorIT} where the prefactor is $E$. In the semiclassical limit, the leading contributions to the vacuum non-persistence and the current are the same. In particular, we can compare \eqref{eq: semiclass} to the scalar current
\begin{equation}\label{eq: current}
    J= \eta\; \frac{\sinh 2\pi E}{\sinh 2\pi \eta}
\end{equation}
for a local inertial observer in the planar patch in the global vacuum \cite{Frob:2014zka}. It equals the current seen by an observer on the static patch wordline, taking into account that the global vacuum appears thermal to such an observer. The leading semiclassical terms in \eqref{eq: semiclass} and \eqref{eq: current} indeed take the same form. Notably, the thermal vacuum non-persistence, just like the current, has a non-zero first derivative\footnote{One might imagine that an already present thermal density of particles is set in motion when turning on the field, but it turns out that such a picture is not sufficient to explain the low $E$ behavior \cite{Frob:2014zka}.} as function of the electric field at $E=0$. The non-thermal result instead has vanishing derivative, as can be seen in Fig. \ref{fig:imbehaviorscal}.

%%%%%%%%%%%%%%%%%%%%%%%%%%%%%%%%%%%%%
\subsection{Revisiting \texorpdfstring{$AdS_2$}{} pair production via the character integral}
%%%%%%%%%%%%%%%%%%%%%%%%%%%%%%%%%%%%%
Our discussion so far has been limited to $dS$, but in fact it was originally noted by \cite{Keeler:2014hba} in the context of $AdS$ that the character encodes both the continuous spectrum eigenfunctions appearing in the heat kernel and the discrete spectrum of quasinormal modes. In \cite{Sun:2020ame}, the Euclidean partition function on the Poincar\'e disk in general dimensions was interpreted in terms of a quasicanonical thermal partition function in Rindler-$AdS$, analogous to the thermal static patch interpretation of the $dS$ results by \cite{Anninos:2020hfj} on which the previous sections were based. Here, we generalize the derivation in \cite{Sun:2020ame} to include a background electric field $E$ in $AdS_2$. This arises naturally in the near-horizon limit of extremal Kerr black holes \cite{Anninos:2019oka}.

As before, we start from the Euclidean result. In particular, we start from the $\mathbb{H}^2$ density of states for a charged scalar in a background field and rotate $B\to iE$ to obtain \cite{Pioline:2005pf,Comtet:1984mm}
\begin{equation}\label{eq: densityads}
  \rho(\lambda) =  \frac{V}{2\pi}\lambda\frac{\sinh2\pi\lambda}{\cosh2\pi\lambda + \cosh2\pi E}\;,
\end{equation}
where $V$ is the regularized volume of $\mathbb{H}^2$.
The 1-loop partition function can be written as
\begin{equation}\label{eq: ads2scal}
    \log Z = - \int^\infty_0 \frac{d\tau}{2\tau}e^{-\frac{\epsilon^2}{4\tau}}\int^{\infty}_{-\infty}d\lambda \rho(\lambda)e^{-\tau(\lambda^2 + \tilde{\eta}_\phi^2)}\;,
\end{equation}
where $\tilde{\eta}_\phi^2 = m^2 +1/4 - E^2$. Similar to the static patch story from before, the thermal nature of the $AdS_2$ black hole \cite{Keeler:2014hba} is taken into account here because the calculation of \eqref{eq: densityads} relied on finding suitably normalized modes on the Poincar\'e disk \cite{Comtet:1984mm, Camporesi:1994ga}, for which the Euclidean time of the black hole is an angular coordinate. 

Inspired by Sec. 5 of \cite{Sun:2020ame}, we define the combination
\begin{equation}
    W(t) = - \frac{e^{-t/2}}{1-e^{-t}}\Big(\frac{1+e^{-t}}{1-e^{-t}}\cos Et +2 E \sin Et\Big)\;,
\end{equation}
to represent the density of states \eqref{eq: densityads}, with $0< \delta < \epsilon$, as
\begin{equation}
 \rho(\lambda) = \frac{V}{4\pi}\Big(\int_{\mathbb{R}+i\delta}+\int_{\mathbb{R}-i\delta}\Big)\frac{dt}{2\pi} W(t) e^{-i\lambda t}\;.
\end{equation}
Plugging this back in \eqref{eq: ads2scal}, we find the regularized expression
\begin{equation}
    \log Z = - \frac{V}{4\pi} \int_{\mathbb{R}+i\delta} \frac{dt}{\sqrt{t^2 +\epsilon^2}}e^{-\tilde{\eta}_\phi\sqrt{t^2 +\epsilon^2}}W(t)\;.
\end{equation}
Formally taking $\epsilon \to 0$, we thus obtain the character integral
\begin{equation} \label{adsscalarlogZ}
 \boxed{  \log Z = \frac{V}{2\pi} \int^\infty_0 \frac{dt}{t}\Big(\frac{1+e^{-t}}{1-e^{-t}}\cos Et + 2 E \sin Et\Big)\chi_{AdS}(t)\;,\quad \chi_{AdS}(t) \equiv \frac{e^{-(\frac12 + \tilde{\eta}_\phi)t}}{1-e^{-t}}\;.\;}
\end{equation}
Following \cite{Sun:2020ame}, this can be evaluated exactly in heat kernel regularization to give
\begin{equation}
\begin{split}
    \log Z = &- \frac{1}{\epsilon^2}-\frac{1}{2} \zeta'(-1,\Delta+iE) - \frac{1}{2}\zeta'(-1, \Delta - iE) + \frac{\tilde{\eta}_\phi}{2} \zeta'(0,\Delta+iE)  \\&+ \frac{\tilde{\eta}_\phi}{2}\zeta'(0,\Delta-iE) - \frac{\tilde{\eta}_\phi^2}{2} + \Big(\frac{1}{24}+E-\frac{E^2}{2}+\frac{\tilde{\eta}_\phi^2}{2}\Big) \log M  \,,
\end{split}
\end{equation}
where $M \equiv 2 e^{-\gamma}/\epsilon$ as before.

Note that for $E$ below the threshold $\sqrt{m^2+ 1/4}$, $\tilde{\eta}_\phi$ is real and \eqref{adsscalarlogZ} takes the exact same form as the real part of the $dS$ character integral \eqref{eq:scalarchar}, with $\chi_{dS}$ replaced by $\chi_{AdS}$. However, for $E$ above the threshold, the character integral becomes complex. One finds
\begin{equation}\begin{split}\label{eq: imAdS}
  \text{Im}[\log Z] = & \; \frac{V|\tilde{\eta}_\phi|}{4\pi}\Big(\text{Li}_1(-e^{-2\pi(E+|\tilde{\eta}_\phi|)})+\text{Li}_1(-e^{-2\pi(E-|\tilde{\eta}_\phi|)})\Big)\\
   &+\frac{V}{8\pi^2}\Big(\text{Li}_2(-e^{-2\pi(E+|\tilde{\eta}_\phi|)})-\text{Li}_2(-e^{-2\pi(E-|\tilde{\eta}_\phi|)})\Big) \;,\end{split}
\end{equation}
in agreement with the result obtained by $\zeta$-function regularization in App. A of \cite{Pioline:2005pf}, which differs from their main result\footnote{
This result is not derived through $\zeta$-function regularization but by rotating $(a,b)\to e^{i\pi}(a,b)$ parameters of the Bessel type integral (3.21) in \cite{Pioline:2005pf}. It should be noted that if one rotates $(a,b)\to (e^{i\pi}a,\frac12(e^{i\pi}+e^{-i\pi})b)$ instead, one arrives at the $\zeta$-function regularized result \eqref{eq: imAdS} by a shortcut.}.  Indeed, in their main result, the vacuum persistence is discontinuous at threshold, whereas \eqref{eq: imAdS} vanishes when $\tilde{\eta}_\phi=0$, which was seen as an indication that BPS particles are not emitted in this vacuum \cite{Pioline:2005pf}. This is also the main difference with $dS$, which has no threshold, as can be understood by the fact that the gravitational field helps pull particles out of the vacuum, while it acts as a potential well in $AdS$. Even if we consider $\tilde{\eta}_\phi=0$ in the scalar case in $dS$, pair production does not vanish unless also $E=0$. 

Since the uniform electric field gives a constant acceleration $a$ to test particles, the existence of a threshold in $AdS$ is related to the fact that one needs a minimum $a$ as compared to the $AdS$ length in order for the Unruh temperature to be well-defined \cite{Deser:1998xb}. It is also worth noting that $\chi_{AdS}$ contains half of the $SO(1,2)$ principal series character, since only the $\Delta$-mode is dynamical. Combining with its shadow $\bar{\Delta}$, we would get the entire $SO(1,2)$ character. In that case, the imaginary parts in the partition function cancel out against each other, suggesting a more stable setup.

Similarly, to find the partition function for a spinor in a background field in $AdS_2$, we start by rotating $B\rightarrow iE$ in the $\mathbb{H}^2$ density of states \cite{Pioline:2005pf,Comtet:1984mm}:
\begin{equation}\label{eq: densityspinads}
    \rho(\lambda) = \frac{V}{\pi} \lambda \frac{\sinh 2 \pi \lambda}{\cosh 2 \pi \lambda - \cosh 2 \pi E} \,.
\end{equation}
The 1-loop partition function can be written as
\begin{equation}
    \log Z = \int^\infty_0 \frac{d\tau}{2\tau}e^{-\frac{\epsilon^2}{4\tau}}\int^{\infty}_{-\infty}d\lambda \rho(\lambda)e^{-\tau(\lambda^2 + \tilde{\eta}_\psi^2)}\;,
\end{equation}
where $\tilde{\eta}_\psi^2 = m^2 - E^2$. Similar to before, we define the combination
\begin{equation}
    W(t) = -\csch \frac{t}{2} \cos (E t) - 2 E \cosh \frac{t}{2} \sin (E t)\;,
\end{equation}
to get the integral representation for the density of states
\begin{equation}
  \rho(\lambda) = \frac{V}{2\pi}\Big(\int_{\mathbb{R}+i\delta}+\int_{\mathbb{R}-i\delta}\Big)\frac{dt}{2\pi} W(t) e^{-i\lambda t}\;,
\end{equation}
with $0 < \delta< \epsilon$. This leads to the fully regularized result
\begin{equation}
  \log Z =  \frac{V}{2\pi}\int_{\mathbb{R}+i\delta} \frac{dt}{\sqrt{t^2+\epsilon^2 }}e^{-\tilde{\eta}_\psi \sqrt{t^2+\epsilon^2}}W(t) \;,
\end{equation}
Formally taking $\epsilon \rightarrow 0$, we arrive at the spinor character integral
\begin{equation}
 \boxed{   \log Z = -\frac{V}{\pi} \int_0^\infty dt \Big( \csch \frac{t}{2} \cos (E t) + 2 E \cosh \frac{t}{2} \sin (E t) \Big) \chi_{AdS}(t) \,, \quad \chi_{AdS}(t) \equiv \frac{e^{-(\frac{1}{2}+\tilde{\eta}_\psi)t}}{1-e^{-t}} \,.}
\end{equation}
Using heat kernel regularization, this partition function can be evaluated exactly to give
\begin{equation}
\begin{split}
    \log Z = \ &\frac{1}{\epsilon^2} +\frac{1}{2} \zeta'(-1,1+\tilde{\eta}_\psi+iE) + \frac{1}{2}\zeta'(-1, 1+\tilde{\eta}_\psi - iE) - \frac{\tilde{\eta}_\psi}{2} \zeta'(0,1+\tilde{\eta}_\psi+iE) \\&- \frac{\tilde{\eta}_\psi}{2}\zeta'(0,1+\tilde{\eta}_\psi-iE) -i \frac{E}{4} \log \frac{\tilde{\eta}_\psi+i E}{\tilde{\eta}_\psi-i E} + \frac{\tilde{\eta}_\psi^2}{2} + \Big(\frac{1}{12}-\frac{E^2}{2}-\frac{\tilde{\eta}_\psi^2}{2}\Big) \log M  \,.
\end{split}
\end{equation}
The partition function becomes complex above the threshold $E>m$, and the imaginary part is given by
\begin{equation}
    \begin{split}
    \text{Im} [\log Z] = &- \frac{V|\tilde{\eta}_\psi|}{2 \pi} \Big( \text{Li}_1(e^{-2 \pi (E+|\tilde{\eta}_\psi|)})+\text{Li}_1(e^{-2\pi(E-|\tilde{\eta}_\psi|)}) \Big) \\
    &-\frac{V}{4\pi^2}\Big( \text{Li}_2(e^{-2\pi(E+|\tilde{\eta}_\psi|)}) + \text{Li}_2(e^{-2\pi(E-|\tilde{\eta}_\psi|)}) \Big) \,,
    \end{split}
\end{equation}
again in agreement with the $\zeta$-function regularized result in App. A of \cite{Pioline:2005pf}.

%%%%%%%%%%%%%%%%%%%%%%%%%%%%%%%%%%%%%
\subsection{Dimensional reduction from Nariai spacetimes}\label{sec: dimred}
%%%%%%%%%%%%%%%%%%%%%%%%%%%%%%%%%%%%%

 In this final and more speculative section, we discuss how our results relate to higher dimensional setups and recently proposed constraints on EFT in $dS$. Charged particles in $dS_2$ can arise from scalars or spinors in a charged Nariai spacetime, as in \cite{Belgiorno:2009pq, Belgiorno:2009da}. This spacetime is $dS_2 \times S^2$, with metric and field strength
\begin{equation}
    ds^2 = -\Big(1-\frac{r^2}{l^2}\Big)dt^2 + \Big(1-\frac{r^2}{l^2}\Big)^{-1}d r^2+ r^2_c d\Omega^2\;,\quad F = \frac{g^2 Q}{4\pi r^2_c}d r\wedge dt\;.
\end{equation}
The 4D $U(1)$-coupling $g$ is kept explicit and the radius $r_c$ of the $S^2$ is determined by the charge $Q$, see \cite{Montero:2019ekk, Montero:2021otb}. The electric field on $dS_2$ is indeed constant. After dimensional reduction, the 2D coupling $e$ satisfies $g^2 = 4\pi r^2_c e^2$. A charged scalar of mass $m$ in 4D gives rise to a KK tower of states with 2D masses 
\begin{equation}\label{eq: kkmasses}
    m^2_2 = m^2 + \frac{n(n+1)}{r^2_c}\;.
\end{equation}

Based on the Swampland-style demand that Nariai solutions should evaporate smoothly to empty de Sitter space, it has been proposed that one should require \cite{Montero:2019ekk} 
\begin{equation}\label{eq: FL}
    m^4 \geq 2 g^2 V =  \frac{48 \pi g^2}{Gl^2_4}\;,
\end{equation}
where in the last step the vacuum energy $V$ was rewritten in terms of the 4D Newton constant $G$ and de Sitter length $l_4$. The proposal \eqref{eq: FL} is known as the FL bound. Moreover, in de Sitter, one has the following bound on the $U(1)$ coupling \cite{Huang:2006hc}
\begin{equation}
    g^2 \geq  \frac{3G}{16\pi l^2_4}\;.
\end{equation}
Substituting this back in \eqref{eq: FL}, one finds the bound (2.31) from \cite{Montero:2021otb}, namely 
\begin{equation}\label{eq: bound}
m^2 \geq \frac{3}{l^2_4}= \frac{1}{l^2}\;,
\end{equation}
where $l$ is the 2D de Sitter length. This means that $\eta>0$ always, and in particular one never reaches the IR hyperconductivity region noted in Sec. \ref{sec:compnonthermal} as well as \cite{Frob:2014zka}, and \cite{Kobayashi:2014zza}, where it was used to derive constraints on magnetogenesis in the early universe. Similarly, the parameter range where energy and entropy are non-monotonic (see Fig. \ref{fig:thermo}) is excluded. In a certain sense, \eqref{eq: bound} corresponds to the lowest energy that can be measured in de Sitter. Modes with a smaller effective mass are frozen by Hubble friction \cite{Montero:2021otb}.

A similar analysis can be done for the rotating Nariai spacetime. As noted in \cite{Anninos:2010gh}, the radial equation for a scalar in this background reduces to that of a charged scalar in $dS_2$ in a constant electric field, where charge and field strength are determined by the angular momentum quantum number on $S^2$ and the rotational parameter of the Nariai spacetime. Schwinger pair production then acts to reduce the angular momentum of the spacetime.

Because of symmetry reasons, the higher-dimensional application of our results is basically restricted to Nariai-like spacetimes. In the $dS_4$ static patch, having an electric field strength would break de Sitter isometries. Moreover, on the Euclidean side, since $\pi_2(S^4) = 0$, there are no nontrivial 2-cycles to wrap flux on. To keep the symmetry intact, one could consider a 4-form proportional to the volume form, since $\pi_4(S^4)= \mathbb{Z}$. This 4-form field strength couples to $2$-branes, for which the semiclassical analysis was done in \cite{Garriga:1994bm}.

%%%%%%%%%%%%%%%%%%%%%%%%%%%%%%%%%%%%%
%%%%%%%%%%%%%%%%%%%%%%%%%%%%%%%%%%%%%
\section{Conclusion}
%%%%%%%%%%%%%%%%%%%%%%%%%%%%%%%%%%%%%
%%%%%%%%%%%%%%%%%%%%%%%%%%%%%%%%%%%%%
In this work, we explored various aspects of static patch physics in a uniform electric field. To do so, we derived a character integral representation of the 1-loop partition function for charged scalars \eqref{eq:scalarchar} and spinors \eqref{eq:spinorchar} in the presence of the background $U(1)$ field on $S^2$, and found the exact results \eqref{eq:scalar_fin} and \eqref{eq:spinor_fin} in terms of Hurwitz $\zeta$-functions. As in \cite{Anninos:2020hfj}, the sphere partition function had a quasicanonical interpretation in the $dS_2$ static patch. The character was obtained both directly as a trace over boundary states and as a sum over quasinormal modes. The isometry group is still $SO(1,2)$, but the generators are shifted by the flux, which also seems to break the extended $SL(2,\mathbb{R})\times SL(2,\mathbb{R})$ symmetry \cite{Anninos:2011af}, as a consequence of which the retarded propagator \eqref{eq: retarded} no longer factorizes. 

An electric field leaves the vacuum subject to pair creation, and we calculated the scalar \eqref{eq:imscalar} and spinor \eqref{eq:imspinor} vacuum persistence for the $dS_2$ static patch by analytic continuation from $S^2$. Keeping Euclidean time periodic, we included the effect of the static patch temperature, which enhances the pair creation rate. This is the main difference compared to \cite{Belgiorno:2009pq,Belgiorno:2009da}. Further arguments came from comparing our results to the current $J$ created in the $dS_2$ planar patch in the global vacuum \cite{Frob:2014zka}, which is thermal to a static patch observer. The leading semiclassical contributions \eqref{eq: semiclass} due the (anti)screening instantons \cite{Garriga:1993fh, Garriga:1994bm} indeed appear with the same prefactor and relative sign in $J$, in contrast to the non-thermal results. Similarly, we find the IR hyperconductity seen in \cite{Frob:2014zka}. Considering dimensional reduction from charged or rotating Nariai black holes \cite{Anninos:2010gh}, the parameter range relevant for hyperconductivity is excluded by the FL bound \cite{Montero:2019ekk, Montero:2021otb}. Finally, applying the character formalism to the $AdS_2$ black hole, we obtained the vacuum persistence \eqref{eq: imAdS} and clarified the assumption on the vacuum implicit in \cite{Pioline:2005pf}. Unlike in $dS_2$, there is a threshold for pair creation in $AdS_2$. 

Below, we suggest some generalizations of our work and list a few ideas for future research:

\begin{itemize}
\item It would be worthwhile to understand if there is a reason why the thermal and background field contributions appear quite clearly separated in the character formulas \eqref{eq:scalarchar} and \eqref{eq:spinorchar}. Likewise, it is unclear to us how the thermally modified transmission coefficients of \cite{Belgiorno:2009pq, Belgiorno:2009da} enter the story.
\item A possible extension is to include the backreaction of the pair creation, both on the electric field and the metric, along the lines of \cite{Brown:1988kg}, by applying a version of the character formalism to Jackiw-Teitelboim gravity. 
\item A surprising feature was the non-monotonicity of the scalar entanglement entropy, see Fig. \ref{fig:thermo}, in the region disallowed by the FL bound \cite{Montero:2019ekk}. It would be good to understand whether the bound is related to general properties of entanglement entropy. Such a connection was noted for the weak gravity conjecture in \cite{Montero:2018fns}.
\item In the spirit of Schwinger's original work \cite{Schwinger:1951nm}, it would be good to flesh out the connection to the worldline formalism, keeping in mind a possible generalization to strings coupled to a top-form in 2+1 dimensions. 
\item One could moreover imagine defining a Nariai character by summing over \eqref{eq: kkmasses}. A similar idea relates the character of conformal fields on the hyperbolic cylinder to characters on the sphere \cite{David:2021wrw, Mukherjee:2021alj}. The relation between partition functions in black hole backgrounds and quasinormal modes is in fact quite general \cite{Denef:2009kn}. Further developing a character formalism in this context would allow for a more elegant definition for the regularized density of states and avoid imposing a brick wall \cite{albert}. 
\item Given the physical relevance of the imaginary part of the partition function, it would be good to understand the role played by the Polchinski phases of \cite{Anninos:2020hfj}. Another gravitational analog consists of coupling 2D quantum gravity with positive cosmological constant to conformal matter with large and positive central charge. Complex saddles besides $dS_2$ will generically make the Euclidean path integral complex valued \cite{Anninos:2021ene}.
\end{itemize}

%%%%%%%%%%%%%%%%%%%%%%%%%%%%%%%%%%%%%
%%%%%%%%%%%%%%%%%%%%%%%%%%%%%%%%%%%%%
\subsection*{Acknowledgements}
%%%%%%%%%%%%%%%%%%%%%%%%%%%%%%%%%%%%%
%%%%%%%%%%%%%%%%%%%%%%%%%%%%%%%%%%%%%
We would like to thank Dionysios Anninos, Frederik Denef, Victor Gorbenko, Albert Law, Gui Pimentel, and Zimo Sun for stimulating discussions, interesting suggestions, and useful comments. MG and KP were supported in part by the U.S. Department of Energy grant de-sc0011941.

\newpage
\appendix
\section{Spherical modes}\label{app:A}
\subsection{Scalar modes}\label{app:A1}
In this appendix, we review the derivation of the scalar spectrum on $S^2$ with a background flux. In what follows, we will take the flux $B$ to be fixed and set $l=1$, restoring it when needed by dimensional analysis. We will also work primarily in the northern coordinate chart $U_N$, although the work can be modified easily to obtain entirely parallel results in the southern coordinate chart $U_S$ (recall that $2B \in \mathbb{Z}$ in order to have smooth gauge transformations between the northern and southern coordinate charts, see \eqref{eq:wu}). Let us consider the spectrum of the magnetic Laplace operator on $S^2$, following \cite{Wu:1976ge}. The action for a charged scalar field in the background produced by the Wu-Yang potential is
\begin{equation}
	S_E = \int d^2x \sqrt{g} \ \phi^* [-(\vec{\nabla} - i \vec{A})^2 +m^2] \phi \,,
\end{equation}
where $A$ is given by \eqref{eq:wu}. To compute the 1-loop partition function of this charged scalar, we must first determine the spectrum of the differential operator $D^2\equiv -(\vec{\nabla}-i\vec{A})^2$. Expanding the operator, we see that
\begin{equation}\begin{split}
	D^2\phi &= -\nabla^2 \phi + i A^\mu \partial_\mu \phi + A^2 \phi + \frac{i}{\sqrt{g}} \partial_\mu (\sqrt{g} A^\mu \phi)  \\
	&=-\nabla^2 \phi + 2i A^\varphi \partial_\varphi \phi + A^2 \phi \,,
\end{split}\end{equation}
where $\nabla^2$ is the Laplace operator on $S^2$:% of radius $l$
\begin{equation}
	\nabla^2 = \frac{1}{\sin \theta}\partial_\theta \sin \theta \partial_\theta + \frac{1}{\sin^2 \theta} \partial_\varphi^2 \,.
\end{equation}
Thus, we find
\begin{equation}
	D^2\phi = -\Big( \frac{1}{\sin \theta} \partial_\theta \sin \theta \partial_\theta  + \frac{1}{\sin^2 \theta} \partial_\varphi^2  - \frac{i2B(1-\cos\theta)}{\sin^2 \theta} \partial_\varphi - \frac{B^2 (1-\cos\theta)^2}{ \sin^2 \theta} \Big) \phi = \lambda \phi \,.
\end{equation}
We now review how the spectrum of this operator was obtained by Wu and Yang \cite{Wu:1976ge} by constructing an $\mathfrak{so}(3)$ algebra. 

Working in the northern coordinate chart, consider the operators
\begin{equation}\begin{split}	\label{eq:scalar_angular}
	L_z &= -i\partial_\varphi - B \,,  \\
	L_{\pm} &= e^{\pm i \varphi} \Big( \pm \partial_\theta + i\cot \theta \partial_\varphi - B\frac{1-\cos \theta}{\sin \theta} \Big)\,.
\end{split}\end{equation}
The quadratic Casimir of this algebra is given by
\begin{equation}\begin{split} 	\label{eq:scalarcasimir}
	L^2 &= L_z^2 + \frac{1}{2}(L_+L_- + L_-L_+)  \\
	&= D^2 + B^2 \,,
\end{split}\end{equation}
with eigenvalues $\ell(\ell+1)$, where $\ell$ can be a non-negative integer or half-integer. By direct computation, it can be verified that these operators satisfy the usual $\mathfrak{so}(3)$ commutation relations
\begin{equation} \label{eq:shifted angular}
	[L_z, L_{\pm}] = \pm L_\pm \,, \quad [L_+,L_-] = 2L_z \,, \quad [L^2,L_z] = [L^2,L_\pm] = 0 \,.
\end{equation}
From the usual analysis, we find a degeneracy of $2\ell+1$ corresponding to the different possible eigenvalues $m$ of $L_z$. Looking at the form of $L_z$ in \eqref{eq:scalar_angular}, we see that the $\varphi$-dependence of the scalar eigenfunctions $Y_{B,\ell,m}$ takes the form\footnote{We follow the notation of \cite{Wu:1976ge}, in which these eigenfunctions were referred to as monopole harmonics.}
\begin{equation}
	Y_{B,\ell,m}(\theta,\varphi) = \Theta_{B,\ell,m}(\theta) e^{i(m+B)\varphi} \,.
\end{equation}
In order for the eigenfunctions to be single-valued, there is the additional constraint that $m +B$ must be an integer, which therefore implies that $\ell+N/2$ must also be an integer. Combining this with \eqref{eq:scalarcasimir}, we may write $\ell$ as\footnote{Here we also used the fact that $2B$ is an integer to obtain the more general result $\ell\pm B \in \mathbb{Z}$. Alternatively, one can analyze the angular momentum operators in the southern coordinate chart.}
\begin{equation}
	\ell = n + |B| \,, \quad n = 0,1,2,\ldots \,.
\end{equation}
Thus, from \eqref{eq:scalarcasimir}, the eigenvalues of $D^2$ are given by
\begin{equation}\label{appA1:eig}
	\lambda_{B,n} =  \ell(\ell+1) - B^2 = (n+|B|)(n+|B|+1) - B^2\,,
\end{equation}
with degeneracy
\begin{equation}\label{appA1:deg} 
	D_{n + B}^{3} = 2\ell+1 = 2(n+|B|) + 1 \,.
\end{equation}

\subsection{Spinor modes}\label{app:A2}
Inspired by the Wu-Yang solution for the scalar case, we can try to perform a similar procedure in the spinor case. Consider a Dirac spinor on $S^2$ in the Wu-Yang background:
\begin{equation}
		S_E =  \int d^2x \sqrt{g} \ \psibar [(\dslash -i\aslash) + m] \psi \,.
\end{equation}
Introducing a zweibein $\tensor{e}{^a_\mu}$,
\begin{equation}
	\tensor{e}{^0_\mu} = (1,0) \,, \quad \tensor{e}{^1_\mu} = (0,\sin\theta) \,,
\end{equation}
and spin connection $\tensor{\omega}{_\mu^{a b}}$,
\begin{equation}
	\tensor{\omega}{_\mu^{ab}} = \tensor{\Gamma}{^{ab}_\mu} + e^{a\nu}\partial_\mu \tensor{e}{^b_\nu} \,, 
\end{equation}
the covariant derivative is given by
\begin{equation}
	\nabla_\mu \psi = \Big(\partial_\mu + \frac{1}{4}\tensor{\omega}{_\mu^{ab}} \sigma_{ab} \Big) \psi  \,,
\end{equation}
where $\sigma_{ab} \equiv \frac{1}{2}[\gamma_a,\gamma_b]$. In two dimensions, we can take the gamma matrices to be $\gamma^0 = \sigma_x, \gamma^1 = \sigma_y$. The relevant non-vanishing Christoffel symbols are
\begin{equation}\begin{split}	\tensor{\Gamma}{^{01}_\varphi} = -\tensor{\Gamma}{ ^{10}_\varphi} = -\cos \theta \,.
\end{split}\end{equation}
Putting these pieces together, we find
\begin{equation}
	\nabla_\theta = \partial_\theta \,, \quad \nabla_\varphi = \partial_\varphi - \frac{i \sigma_z}{2} \cos\theta \,.
\end{equation}

We are interested in the spectrum of the differential operator $\Dslash \equiv -i(\dslash - i\aslash$), which takes the explicit form
\begin{equation}\begin{split}	\label{eq:grad}
	\Dslash &= -i\gamma^a \tensor{e}{_a^\mu} (\nabla_\mu - iA_\mu)  \\ 
	&= -i \sigma_x \Big(\partial_\theta + \frac{1}{2}\cot \theta\Big) - \sigma_y \Big(\frac{i}{\sin \theta} \partial_\varphi + B\frac{1-\cos\theta}{\sin\theta}\Big) \,.
\end{split}\end{equation}
For future use, we also calculate the square of the operator $\Dslash$ to find
\begin{equation}\begin{split}\label{eq:Dsquared}
	\Dslash^2 &= \Big( - \frac{1}{\sin \theta} \partial_\theta \sin \theta \partial_\theta - \frac{1}{\sin^2\theta} \partial_\phi^2 + \frac{i \sigma_z \cos\theta}{\sin^2 \theta}\partial_\varphi + \frac{1}{4 \sin^2\theta} + \frac{1}{4} \Big) \\
	& \quad \ + \frac{1}{\sin^2\theta} \Big( i2B(1-\cos\theta)\partial_\varphi + B\sigma_z \cos \theta(1-\cos\theta) + B^2(1-\cos\theta)^2  \Big) -B \sigma_z \,.
\end{split}\end{equation}
As before, we construct an $\mathfrak{su}(2)$ algebra, which in the northern coordinate chart, takes the form  \cite{fakhri2007,Abrikosov:2002jr}
\begin{equation}\begin{split}
	J_z &= -i\partial_\varphi - B \,, \\
	J_{\pm} &= e^{\pm i \varphi} \Big( \pm \partial_\theta + i\cot \theta \partial_\varphi + \frac{\sigma_z}{2 \sin \theta} - B\frac{1-\cos \theta}{\sin \theta} \Big)\,.
\end{split}\end{equation}
Direct computation shows that these operators satisfy the usual commutation relations
\begin{equation}
	[J_z, J_{\pm}] = \pm J_\pm \,, \quad [J_+,J_-] = 2J_z \,.
\end{equation}
Furthermore, it can also be shown that $\Dslash$ commutes with the above operators:
\begin{equation}
	[\Dslash, J_z] = [\Dslash, J_+] = [\Dslash, J_-] = 0 \,.
\end{equation}
The quadratic Casimir $J^2$ is given by
\begin{equation}\begin{split}
	J^2 &= J_z^2 + \frac{1}{2}(J_+J_- + J_-J_+)  \\
	&= \Dslash^2 + B^2-\frac{1}{4} \,,
	\label{eq:spinorcasimir}
\end{split}\end{equation}
with eigenvalues $j(j+1)$ for $j$ a non-negative integer or half-integer. Again, from the usual analysis, there is a degeneracy of $2j+1$ corresponding to eigenvalues $m$ of $J_z$. When $J_z$ acts on eigenspinors, denoted $\Upsilon_{B,j,m}$, we have
\begin{equation}
	J_z \Upsilon_{B,j,m} = (-i\partial_\varphi -B) \Upsilon_{N,j,m} = m \Upsilon_{N,j,m} \,.
\end{equation}
Thus, the $\varphi$ dependence of $\Upsilon_{B,j,m}$ takes the form
\begin{equation} \label{eq:eigenspinors}
	\Upsilon_{B,j,m}(\theta,\varphi) = \Xi_{B,j,m}(\theta) e^{i(m+B)\varphi}  \,.
\end{equation}
In order to furnish a spin-$\frac{1}{2}$ $SU(2)$ representation, there is the additional constraint that $m +B$ must be a half-integer, which therefore implies that $j+B$ must also be a half-integer. Combining this with \eqref{eq:spinorcasimir}, we may represent $j$ as
\begin{equation}
	j = n + |B| - \frac{1}{2}  \,, \quad n = 0,1,2,\ldots \,.
\end{equation}
From \eqref{eq:spinorcasimir}, the eigenvalues $\lambda_{B,n}^2$ of $\Dslash^2$ therefore satisfy
\begin{equation}
	\lambda_{B,n}^2 = j(j+1) - B^2 + \frac{1}{4} = n(n + 2|B|) \,,
\end{equation}
with degeneracy
\begin{equation}	\label{eq:dirac_mult}
	D_{n+B-\frac{1}{2}, \frac{1}{2}}^{3} = 2j+1 = 2(n + |B|) \,.
\end{equation}
The eigenvalues of $\Dslash$ are therefore
\begin{equation}
    \lambda_{N,n} = \pm \sqrt{n(n+2|B|)} \,,
\end{equation}
with degeneracy \eqref{eq:dirac_mult}.

The eigenspinors\footnote{We are slightly changing notation from $\Upsilon_{B,j,m} \rightarrow\Upsilon_{B,n,m}$ for simplicity.} $\Upsilon_{B,n,m}$ of $\Dslash^2$ can be constructed explicitly by first noting that \eqref{eq:Dsquared} does not mix the spinor components, so we can individually solve for the upper and lower components. We consider the case $B>0$ in the following. Note that if $\Upsilon_{B,n,m}(\theta,\varphi)$ is an eigenspinor with $B > 0$,
\begin{equation} \label{eq:positiveNspinor}
    \Dslash \Upsilon_{B,n,m}(\theta,\varphi) = \lambda_{B,n} \Upsilon_{B,n,m}(\theta,\varphi)\,,
\end{equation}
then $\sigma_x \Upsilon_{-B,n,m}(\theta,-\varphi)$ (equivalently $\sigma_x \Upsilon_{-B,n,-m}(\theta,\varphi)$) is the eigenspinor corresponding to $B < 0$. To see this, note that multiplying \eqref{eq:grad} on the left and right with $\sigma_x$ and taking $\varphi \rightarrow - \varphi$ effectively takes $B \rightarrow -B$. Therefore,
\begin{equation}
    \Big( \sigma_x \Dslash \sigma_x \Big)   \sigma_x \Upsilon_{-B,n,m}(\theta, -\varphi) = \lambda_{B,n} \sigma_x \Upsilon_{-B,n,m}(\theta,-\varphi)\,,
\end{equation}
so the $B<0$ solutions can be obtained from the $B>0$ ones.

Focusing on the $\theta$-dependent piece $\Xi_{B,n,m}(\theta)$, for the upper component $\psi(\theta)$, we find the independent solutions
\begin{equation}\begin{split}
\psi^\pm_{B,n,m}(\theta) =\; &(1-\cos\theta)^{\pm \frac14(-1+2m+2B)}(1+\cos\theta)^{\pm \frac14(-1-2m+2B)} 
\\ &\times\; {_2F_1}\Big(\frac{1}{2}\mp \big(n + \frac{1}{2}\big), \frac{1}{2} \pm \big( n+2B -\frac{1}{2} \big); 1 \pm \big(m+B-\frac{1}{2}\big); \sin^2{\frac{\theta}{2}}\Big) \,. \end{split}
\end{equation}
Regularity on $\theta \in [0, \pi]$ requires $m + B+1/2 \in \mathbb{Z}$. For $2m + 2B+1 > 0$, $\psi^+$ is regular. For more negative values of $m$ (which must still satisfy $|m|\leq j = n +|B|-\frac{1}{2} $), the solution $\psi^-$ is the regular one. The same can be done for the lower component $\phi(\theta)$, with the result
\begin{equation}\begin{split}
    \phi^\pm_{B,n,m}(\theta)= \;& (1 -\cos\theta)^{\pm \frac14(1 + 2 m + 2B)} (1 + \cos\theta)^{\pm \frac14 (1 - 2 m + 2B)}
    \\&\times \;  {_2F_1}\Big(\frac{1}{2} \mp \big(n-\frac{1}{2}\big), \frac{1}{2} \pm \big(n+2B+\frac{1}{2}\big); 1 \pm \big(m+B + \frac{1}{2}\big); \sin^2{\frac{\theta}{2}}\Big) \,. \end{split}
\end{equation}
These modes have eigenvalues $\lambda_{B,n}$. 

We can now combine the above modes to obtain the eigenspinors of $\Dslash$. For $2m + 2B+1 > 0$ the regular eigenmodes are 
\begin{equation}
    \Xi^{\pm}_{B,n,m} = \Big(\psi^+_{B,n,m},  \pm \frac{ i\sqrt{n(n+2B)}}{1+2m+2B}\phi^+_{B,n,m}\Big)^T \,,
\end{equation}
with eigenvalues $\pm \sqrt{n(n+2B)}$. For $2m + 2B+1 \leq 0$, the regular eigenmodes are
\begin{equation}
    \Xi^\pm_{B,n,m} = \Big(\pm \frac{i\sqrt{n(n+2B)}}{1-2m-2B}\psi^-_{B,n,m}, \phi^-_{B,n,m}\Big)^T \,,
\end{equation}
also with eigenvalues $\pm \sqrt{n(n+2B)}$. In particular, note that there are $2B$ zeromodes of $\Dslash$ corresponding to $n=0$:
\begin{equation}
    \Xi_{N,0,m}(\theta) = \Big( (1-\cos \theta)^{\frac{1}{4}(-1+2m+2B)}(1+\cos \theta)^{\frac{1}{4}(-1-2m+2B)},0\Big)^T \,.
\end{equation}

%%%%%%%%%%%%%%%%%%%%%%%%%%%
\section{\texorpdfstring{$\Im [\log Z_\phi]$}{} from the Green function}\label{app: B}
%%%%%%%%%%%%%%%%%%%%%%%%%%%

In this appendix, we verify the imaginary part of the consistency relation \eqref{eq: logZvsGF}, providing an alternative to the heat kernel derivation of $\text{Im}[\log Z_\phi]$ in Sec. \ref{sec: schwingscal}. Using the coincident limit of the Green function in \eqref{Wightman_solved},
\begin{equation}
    G(z \rightarrow 1) \approx -\frac{1}{4\pi} \Big( 2 \gamma + \psi(\Delta+ i E) + \psi(\bar{\Delta}+iE ) + \log(1-z)\Big) \,,
\end{equation}
we want to verify the consistency relation
\begin{equation}
    \Im[\log Z_\phi] = \frac{V}{4 \pi} \Im\Big[ \int^{m^2} d\bar{m}^2 \ \Big(\psi(\Delta + iE) + \psi(\bar{\Delta} + i E)\Big)\Big]\,.
\end{equation}
Converting the integral over $\bar{m}^2$ to an integral over $\Delta$ gives the result
\begin{equation}\begin{split}
    \Im[\log Z_\phi] = \frac{V}{2 \pi} \text{Im}\Big[&i \eta \Big(\log \Gamma(\bar{\Delta} + i E ) - \log \Gamma(\Delta + i E )\Big)  \\
    &+ \psi^{(-2)}(\bar{\Delta} + i E) + \psi^{(-2)}(\Delta + i E) \Big] \,.
\end{split}\end{equation}
Now using the relations \cite{ADAMCHIK1998191} 
\begin{equation}\begin{split}
    \zeta'(0,a) &= \log \Gamma(a) - \frac{1}{2}\log 2 \pi \,,  \\
    \psi^{(-2)}(z) &= \frac{(1-z)z}{2} + \frac{z}{2} \log 2 \pi - \zeta'(-1) + \zeta'(-1,z)  \,,
\end{split}\end{equation}
and dropping the real terms, we find
\begin{equation}\begin{split}
    \Im[\log Z_\phi] = \frac{V}{2 \pi} \text{Im}\Big[ &i \eta \Big(\zeta'(0,(\bar{\Delta} + i E ) - \zeta'(0,(\Delta + i E )\Big)   \\
    &+ \zeta'(-1,\bar{\Delta} + i E) + \zeta'(-1,\Delta + i E) \Big] \,,
\end{split}\end{equation}
which precisely agrees with the imaginary part of the exact one-loop path integral \eqref{eq:scalar_fin}, which can also be written in the form \eqref{eq:imscalar}.

%%%%%%%%%%%%%%%%%%%%%%%%%%%%%%%%%%%%%
% Bibliography
%%%%%%%%%%%%%%%%%%%%%%%%%%%%%%%%%%%%%

\newpage
\setlength{\parskip}{9.7pt}
\bibliographystyle{JHEP.bst}
\bibliography{char2d}

%%%%%%%%%%%%%%%%%%%%%%%%%%%%%%%%%%%%%
	
\end{document}